\documentclass[amsmath,amssymb,showpacs,floatfix,twocolumn,prb,superscriptaddress]{revtex4}
\usepackage{graphicx}
\usepackage{dcolumn}
\usepackage{bm}

\graphicspath{{figures/}}

\def\iomn{i\omega_n}
\def\vk{\mathbf k}
\def\vK{\mathbf K}
\def\ek{\varepsilon_{\mathbf k}}
\def\al{\alpha}
\def\be{\beta}
\def\Ga{\Gamma}

\begin{document}

\title{Pseudogap opening and formation of Fermi arcs\\ as an
orbital-selective Mott transition in momentum space}

\date{\today}

\author{Michel Ferrero}
\affiliation{Centre de Physique Th{\'e}orique,
Ecole Polytechnique, CNRS, 91128 Palaiseau Cedex, France}
\affiliation{Institut de Physique Th{\'e}orique, CEA, IPhT, CNRS, URA 2306,
91191 Gif-sur-Yvette, France}
\author{Pablo S. Cornaglia}
\affiliation{Centro At{\'o}mico Bariloche and Instituto Balseiro,
CNEA, CONICET, 8400 Bariloche, Argentina}
\affiliation{Centre de Physique Th{\'e}orique,
Ecole Polytechnique, CNRS, 91128 Palaiseau Cedex, France}
\author{Lorenzo De Leo}
\affiliation{Centre de Physique Th{\'e}orique,
Ecole Polytechnique, CNRS, 91128 Palaiseau Cedex, France}
\author{Olivier Parcollet}
\affiliation{Institut de Physique Th{\'e}orique, CEA, IPhT, CNRS, URA 2306,
91191 Gif-sur-Yvette, France}
\author{Gabriel Kotliar}
\affiliation{Physics Department and Center for Materials Theory, Rutgers University,
Piscataway NJ 08854, USA}
\author{Antoine Georges}
\affiliation{Centre de Physique Th{\'e}orique,
Ecole Polytechnique, CNRS, 91128 Palaiseau Cedex, France}

\begin{abstract}
We present an approach to the normal state of cuprate superconductors which is
based on a minimal cluster extension of dynamical mean-field theory.
Our approach is based on an effective two-impurity model embedded in a
self-consistent bath. The two degrees of freedom of this effective
model can be associated to the nodal and antinodal regions of momentum space.
We find a metal-insulator transition which is selective in momentum space:
At low doping quasiparticles are destroyed in the antinodal region,
while they remain protected in the nodal region, leading to the formation of apparent Fermi arcs.
We compare our results to tunneling and angular-resolved photoemission experiments on cuprates. 
At very low energy, a simple description of this transition can be given
using rotationally invariant slave bosons.
\end{abstract}


\maketitle

\section{Introduction and motivations}
\label{sec:intro}

The doping of a Mott insulator is a fundamental problem of condensed matter physics,
which has attracted considerable attention in view of its
relevance to the physics of cuprate superconductors.~\cite{Anderson_1987}
In the simplest Brinkman-Rice~\cite{Brinkman1970} description, the doped metallic state
is a Fermi liquid in which quasiparticles are formed with a heavy
mass $m^*/m\sim1/\delta$ and a reduced weight $Z\sim\delta$
($\delta$ is the doping level).
This physical picture can indeed be rationalized using the modern theoretical
framework of dynamical mean-field theory (DMFT).~\cite{georges1996,kotliar_review_rmp_2006,kotliar_dmft_physicstoday}
DMFT, in its single-site version, is applicable when spatial correlations
are weak, which is favored by high dimensionality and strong competing
(e.g. orbital) fluctuations.

In cuprates however, which are two-dimensional materials with low orbital
degeneracy, it was pointed out long ago by Anderson in a seminal
paper~\cite{Anderson_1987} that the antiferromagnetic superexchange ($J$)
plays a key role, leading to strong short-range correlations associated
with singlet formation (valence bonds) between nearest-neighbor lattice sites.
Slave boson mean-field theories~\cite{Baskaran_1987,Kotliar_1988,grilli_prl_1990,Fukuyama88,LeeRMP06}
as well as projected variational wave-functions,~\cite{Gros_1988,Paramekanti_2001}
provide simple theoretical frameworks to
incorporate this effect, modifying the Brinkman-Rice
picture at small doping $\delta\lesssim J/t$ and leading in
particular to a finite effective mass of quasiparticles
$m^*/m\sim 1/(J/t+\delta)$. This is indeed consistent
with observations in cuprates, in which only a moderate
enhancement of the effective mass is observed.

However, both single-site DMFT and simple variational or slave-boson
mean-field theories share a common feature, namely that the characteristic
energy (or temperature) scale below which coherent quasiparticles are formed is
uniform along the Fermi surface, and of order $\delta t$ at small doping levels.
This is clearly inconsistent with experimental observations in
underdoped cuprates. Indeed, these materials are
characterized by a strong differentiation of quasiparticle properties
in momentum space, a phenomenon which is key to their
unusual normal-state properties.
In underdoped cuprates, coherent quasiparticle excitations are suppressed
in the antinodal regions, around momenta ($0,\pi$) and ($\pi,0$)
of the Brillouin zone (BZ), and a pseudogap appears
below a characteristic temperature scale (which decreases as the doping level
is increased). Instead, reasonably coherent quasiparticles are
preserved in the nodal regions around ($\pi/2,\pi/2$).
The signature of this phenomenon in angular-resolved photoemission spectroscopy (ARPES)
is the formation of Fermi `arcs' in the underdoped regime, defined as the regions
of momentum space where the spectral function is intense at low excitation energy
(see e.g. Ref.~\onlinecite{damascelli_rmp_2003} for a review).
Suppression of quasiparticle coherence in the antinodal regions is also
apparent from other spectroscopies, such as electronic Raman
scattering~\cite{Letacon_natphys_2006,venturini_prl_2002,guyard_prb_2008} (with
the B$_{1g}$ and B$_{2g}$ channels associated with antinodes and nodes,
respectively) or quasiparticle interference patterns obtained by scanning
tunneling microscopy.~\cite{Kohsaka_nature_2008}

Momentum-space differentiation and the nodal/antinodal dichotomy is therefore
an outstanding challenge for theories of strongly-correlated electrons.
Various lines of attack to this problem have been taken.
At intermediate and strong coupling, and apart from the extremely low-doping region,
correlation lengths are expected to be short (as also supported by experimental observations).
Hence, it is appropriate in this regime to take into account short-range correlations
within cluster extensions of the DMFT framework. Such investigations have been
quite successful (for reviews,
see e.g. Refs.~\onlinecite{maier_cluster_rmp_2005,kotliar_review_rmp_2006,Tremblay06}).
%
%
%
%
Most studies have considered clusters of at least four sites
(a plaquette)~\cite{LichtensteinKatsnelson_SC_2000,Maier2x2_SC_PRL2000,Jarrell_2x2PhaseDiagram_2001EPL,
parcollet_cdmft_finiteTMott_prl_2004,civelli_breakup_prl_2005,
KyungTremblayPeriodizationG, biroli_clusterlong_prb_2004,HauleAvoidedQCP,Capone_2x2EDCompetitionAFSC,
Haule2x2,Haule2x2OpticsEPL,
Civelli2gaps,GullWernerMillis_PlaquetteEPL2008} and
numerical efforts have been devoted to increasing the cluster size in order to improve
momentum-resolution and to advance toward an understanding of the two-dimensional
case.~\cite{MaierSCLargeCluster}

In this article, we follow a different route, looking for
a description based on the {\it minimal cluster}
able to successfully describe momentum-space differentiation together with Mott physics.~\cite{VBDMFT}
We find that a two-site cluster is sufficient to achieve this goal on a qualitative level,
and to a large extent on a quantitative level, when compared to  larger cluster calculations.
Our approach is based on a division of the BZ into two patches, one containing in
particular the $(\pi/2,\pi/2)$ momentum and the other one in particular the $(\pi,0)$ and
$(0,\pi)$ momenta. A mapping onto a two-impurity Anderson model with
self-consistent hybridization functions is made, following a generalization of the
dynamical cluster-approximation construction (very similar results are actually obtained
within a cellular-DMFT self-consistency condition).
The self-energies associated with the `nodal' and `antinodal' patches are
shown to correspond to the bonding (even) and antibonding (odd) orbitals
of the self-consistent impurity model, respectively.
This allows us to construct a `valence-bond dynamical mean-field theory' (VB-DMFT)~\cite{VBDMFT}
of nodal/antinodal differentiation, in which this phenomenon is associated with the
distinct properties of the orbitals associated with different regions of momentum space.

One of our central results is that, below a critical value of the doping level
$\delta_c\simeq 16\%$, the `nodal' (bonding) orbital remains metallic while the
`antinodal' (antibonding) orbital displays a pseudogap. Correspondingly, the
scattering rate associated with the nodal orbital is suppressed in the
pseudogap state, while the antinodal orbital scattering rate increases as the
doping level is reduced.  Hence, nodal/antinodal differentiation corresponds in
this minimal description to an orbital differentiation, a {\it momentum-space
analogue}~\cite{BiermannDeMediciGeorges_OSMTPRL05} of the orbital-selective
Mott transition which has been extensively discussed recently in the different
context of transition-metal oxides with several active
orbitals.~\cite{anisimov:epjb:2002:1} The suppressed coherence of antinodal
quasiparticles clearly originates, in our description, from Mott physics
affecting antinodal regions in a dramatic manner while nodal regions remain
comparatively protected.

A definite advantage and important motivation for building a minimal
description based on the smallest
possible cluster is to  advance our qualitative understanding.
Since the theory is based on a two-site Anderson model,
results can be interpreted in terms of valence-bond singlet formation
and linked to the well-documented competition between singlet-formation
and individual Kondo screening.~\cite{Jones88,FerreroNutshell}
The two-impurity Kondo (or Anderson) model is the
simplest model which captures this competition.
However our findings show that, in contrast to the two-impurity model with
a fixed hybridization to the conduction-electron bath, the additional
self-consistency of the bath which is central to dynamical mean-field
constructions brings in novel aspects. Indeed, the critical point encountered
at a finite doping $\delta_c$ is found only in the lattice model involving
self-consistent baths, while it is replaced by a crossover in the
non self-consistent two-impurity Anderson model.

This article is organized as follows.
In Sec.~\ref{sec:model}, we specify the two-dimensional Hubbard model under
consideration and describe the BZ patching and VB-DMFT mapping onto
a self-consistent two-orbital model.
Then, we briefly review the two main techniques that we have used for the
solution of this problem: the strong-coupling continuous-time quantum Monte Carlo
algorithm~\cite{CTQMC_Werner,CTQMC_Haule} (Sec.~\ref{sec:ctqmc}) and the (semi-analytical) rotationally-invariant
slave-boson approximation~\cite{Lechermann2007a} (Sec.~\ref{sec:risb}).
Sec.~\ref{sec:osmt} is devoted to a detailed presentation of the
orbital-selective transition and of its physical relevance to
nodal/antinodal differentiation. In particular, the frequency dependence of
the self-energies and spectral functions on the real axis are
presented and interpreted.
In Sec.~\ref{sec:impmodel}, these results are contrasted to the
physics of a two-impurity problem with non self-consistent baths,
in connection with the Kondo to RKKY (singlet) crossover observed
there.
Finally, in Sec.~\ref{arcs}, the issue of momentum-space reconstruction
is considered, and the connection to the formation of `Fermi arcs' is
discussed. For the sake of clarity and completeness, some technical aspects of
our work are discussed more in detail in appendices.

\section{Theoretical framework}
\label{sec:theory}

\subsection{Model and valence-bond dynamical mean-field theory}
\label{sec:model}

We study the Hubbard model on a square lattice, with hopping between nearest-neighbor
$t$ and next-nearest-neighbor sites $t'$. The corresponding Hamiltonian is given by
\begin{align}
H =& \sum_{\vk, \sigma = \uparrow,\downarrow}  \ek c^{\dagger}_{\sigma \vk} c_{\sigma \vk} + U \sum_{i} n_{i\downarrow} n_{i\uparrow}
\\
\ek =& -2t \bigl (\cos(\vk_{x}) + \cos (\vk_{y}) \bigr) -4t' \cos(\vk_{x}) \cos(\vk_{y}).
\end{align}
In the following, we use $U/t=10$ and $t'/t=-0.3$, which are values commonly used for modeling
hole-doped cuprates in a single-band framework.
All energies (and temperatures) are expressed in units of $D=4t=1$, and
the doping is denoted by $\delta$.
We restrict ourselves to paramagnetic normal phases.

\begin{figure}[ht!]
  \includegraphics[width=6cm,clip=true]{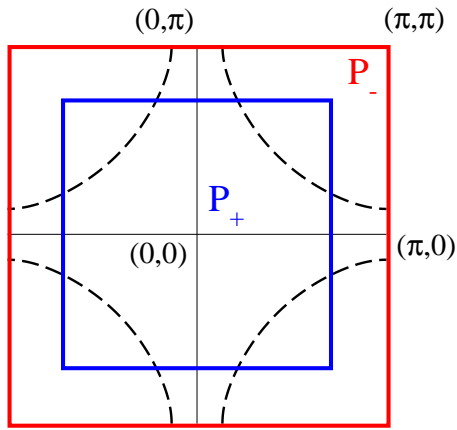}
  \caption{(Color online) The Brillouin zone is divided into
two patches $P_+$ (inside the inner blue square)
and $P_-$ (between the two squares).
The dotted line is the free ($U=0$) Fermi surface at $\delta=0.1$ for $t'/t = -0.3$.
$P_+$ (resp. $P_-$) encloses the nodal (resp. antinodal) region.
}
  \label{fg:patches}
\end{figure}

\begin{figure}[ht!]
  \includegraphics[width=8.5cm,clip=true]{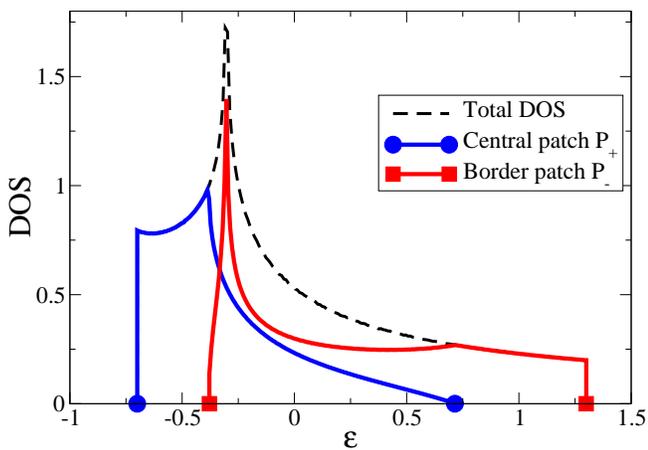}
  \caption{(Color online) Partial density of states of the two patches $P_+$ (solid blue curve with circles)
and $P_-$ (solid red curve with squares), and total density of states (dashed curve); $t'/t = -0.3$.
}
  \label{fg:partial_dos}
\end{figure}

In this paper, we focus on a two-site dynamical cluster approximation (DCA).~\cite{maier_cluster_rmp_2005}
For completeness, we recall the DCA construction in  Appendix~\ref{Appendix-dca}.
The principle of the DCA approximation is to cut the Brillouin zone into
patches and approximate the self-energy as a piecewise constant function on the patches.
The many-body problem can then be solved using an effective self-consistent multiple quantum impurity model.
A priori, there is some arbitrariness in the choice of patches in DCA.
In this paper, we exploit this freedom in order to separate the nodal and the antinodal region
into two patches, so that the properties of each region will be described
by one orbital of the effective impurity model.
More precisely, we choose the minimal set of two patches of equal area
$P_+$ and $P_-$  represented in Fig.~\ref{fg:patches}:
$P_+$ is a central square centered at momentum $(0,0)$ and containing the nodal region;
the complementary region $P_{-}$ extends to the edge of the BZ and contains in particular
the antinodal region and the $(\pi,\pi)$ momentum. On Fig.~\ref{fg:partial_dos}, we also
present the partial density of state of both patches.

It is important to check that the main qualitative results of our
approach are independent of the precise shape of the patches.
We will discuss this point in Sec.~\ref{sec:otherpatches},
and show that indeed our results are qualitatively
similar for a family of patches in which the $P_{+}$ patch
encloses a variable part of the bare Fermi surface
around the nodal point.
Moreover, we have also considered another cluster method, cellular-DMFT
(CDMFT),~\cite{maier_cluster_rmp_2005,kotliar_review_rmp_2006}
and obtained qualitatively similar results. Because two-site CDMFT breaks the lattice square symmetry,
we focus here on a generalized DCA approach.

Following the DCA construction (see also Appendix~\ref{Appendix-dca}), we associate a momentum-independent
self-energy $\Sigma_{\pm}(\omega)$ to each patch of the Brillouin zone.
This self-energy is then identified with the Fourier transform of the cluster self-energy
of a two-site cluster of Anderson impurities embedded in a self-consistent bath.
This two-site Anderson impurity model is given by
\begin{gather}
 S_{\text{eff}} = - \iint_{0}^{\beta } d \tau  d \tau'
\sum_{a,b = 1,2 \atop \sigma = \uparrow,\downarrow}c^{\dagger }_{a \sigma} (\tau )
G_{0, ab}^{-1} (\tau ,\tau')
c^{ }_{b\sigma} (\tau' )\nonumber
\\  + \int_{0}^{\beta } d \tau \label{Seff}
U
\sum_{a = 1,2} n_{a\downarrow} n_{a\uparrow}(\tau)
\\
 G^{-1}_{0ab}(i \omega_n) = (i \omega_n + \mu) \delta_{ab} - \bar t( 1- \delta_{ab}) - \Delta_{ab}(i \omega_{n}),
 %
\end{gather}
where $a,b=1,2$ is the site index, $U$ is the on-site
interaction, $\Delta$ is the hybridization function with a local component $\Delta_{11}(\omega)=\Delta_{22}(\omega)$
and an inter-site one $\Delta_{12}(\omega)$. We choose a convention in which the hybridization $\Delta$
vanishes at infinite frequencies
and therefore denote the constant term separately ($\bar t$).
Since we restrict ourselves to paramagnetic solutions, we dropped the spin dependence
of $G_{0}$, $\Delta$ and $\bar t$.
The self-consistency condition determines both $\Delta$ and $\bar t$ and
is written in the Fourier space of the cluster,
which in this case reduces to the even and odd orbital combinations
$c^{\dagger}_{\pm\sigma} = (c^\dagger_{1\sigma}\pm c^\dagger_{2\sigma})/\sqrt{2}$:
\begin{align}
\Sigma_{K}(i \omega_n) =& G_{0K}(i \omega_n)^{-1} - G_{K}(i \omega_n)^{-1}
\\
G_{K}(i \omega_n)=&\sum_{\vk\in P_{K}} \dfrac{1}{i \omega_n+\mu - \ek - \Sigma_K(i \omega_n) }.
\end{align}
In this expression,  momentum summations are normalized to unity within each patch, and
the index $K=\pm$ refers both to the inner/outer patch index and
to the even/odd orbital combinations of the two-impurity problem.
$\bar t$ is determined by the $1/\omega^2$ expansion of the previous equations, leading to
\begin{equation}
\bar t = \sum_{\vk\in P_{+}} \ek = - \sum_{\vk\in P_{-}} \ek.
\end{equation}
The impurity model has the same local interaction as the original lattice model: This is a consequence of
the fact that both patches have equal surface (see Appendix~\ref{Appendix-dca}).

As usual in the DMFT problems, the quantum impurity model~(\ref{Seff}) can be
rewritten in a Hamiltonian form, i.e. as the Hamiltonian for a
dimer coupled to a self-consistent bath
\begin{equation}
H = H_\mathrm{dimer} + H_\text{bath},
\end{equation}
where $H_\mathrm{dimer}$ can be written in the $1,2$ basis as
\begin{equation}
\label{def:HDimer}
H_\mathrm{dimer} \equiv
\sum_{a,b = 1,2 \atop \sigma = \uparrow,\downarrow}c^{\dagger }_{a \sigma}
\bigl (
\bar t (1-\delta_{ab}) +\varepsilon_0 \delta_{ab}
\bigr )
c^{ }_{b\sigma}
+ \sum_{a = 1,2} U n_{a\downarrow} n_{a\uparrow},
\end{equation}
where $\varepsilon_0 \equiv -\mu$.
Alternatively, $H_\mathrm{dimer}$ can be written in the even/odd basis where the hybridization is diagonal
\begin{multline}
H_\mathrm{dimer} =
\sum_{s = \pm \atop \sigma = \uparrow,\downarrow}c^{\dagger }_{s \sigma}
\left(
s \bar t +\varepsilon_0 
\right)
c^{ }_{s\sigma} + \nonumber \\
 \frac{U}{2} \sum_{s = \pm \atop \bar s = -s}\Bigl(
n_{s\uparrow} n_{s\downarrow}  +  n_{ s \uparrow} n_{\bar s\downarrow}
+  c^{\dagger}_{s\uparrow} c^{\dagger}_{s\downarrow}  c^{}_{\bar s\downarrow} c^{}_{\bar s\uparrow}
+ c^{\dagger}_{s\uparrow} c^{\dagger}_{\bar s\downarrow}  c^{}_{s\downarrow} c^{}_{\bar s\uparrow}
\Bigr).
\end{multline}
Note that, since we will be solving the quantum impurity model
using continuous-time quantum Monte Carlo and rotationally-invariant slave-boson methods,
which work within the action formalism, we will not need the explicit form of the
bath term $H_\text{bath}$.

%

\subsection{Continuous-time Monte Carlo}
\label{sec:ctqmc}

\def\moy#1{\left\langle #1 \right\rangle}
\def\bigmoy#1{\bigl\langle #1 \bigr\rangle}
\def\Bigmoy#1{\Bigl\langle #1 \Bigr\rangle}

A numerically exact solution of the self-consistent two-impurity problem is obtained
using continuous-time quantum Monte Carlo (CTQMC)~\cite{CTQMC_Werner,CTQMC_Haule}
which sums the perturbation theory in 
$\Delta_{ab}(i\omega_n)$ on the Matsubara axis.
The partition function of the impurity model
\[
Z = \int {\cal D} c^{\dagger} {\cal D} c
\exp \bigl ( - S_{\text{eff}} \bigr )
\]
is expanded in powers of the hybridization $\Delta$, leading to
\def\CC{ \prod_{i=1}^{n} c^{\dagger}_{a_{i}\sigma_{i}}(\tau_{i}) c_{a_{i}\sigma_{i}}(\tau'_{i})}
\begin{multline}
Z = \sum_{n\geq 0} \frac{1}{n!}
\int \prod_{i=1}^{n} d\tau_{i} d \tau'_{i}
\sum_{a_{i}=\pm \atop \sigma_{i} = \uparrow,\downarrow}
\det_{1\leq i,j \leq n}
\bigl [
  \Delta_{a_{i}, a_{j}}(\tau_{i} - \tau'_{j})
\bigr]
\times
\\
\mathop{\text{Tr}}
\left( {\cal T}
e^{-\beta H_\mathrm{dimer}} \CC 
\right),
\end{multline}
where ${\cal T}$ is time ordering and $H_{\mathrm{dimer }}$ is given by~(\ref{def:HDimer}).

The partition function is then sampled with the Metropolis algorithm,
where a configuration of size $n$ is given by the $n$ indices $a_{i}, \sigma_{i}$ and times $\tau_{i}$, $\tau_{i}'$
of the $c^{\dagger},c$ operators. The  Monte Carlo probability of a configuration is
given by the absolute value of the product of the trace and the determinant term.
The Green's function is then accumulated using the formula obtained by differentiating $Z$ with respect
to $\Delta$.~\cite{CTQMC_Werner,CTQMC_Haule}

Another quantity of interest is the relative weight of multiplets of the dimer problem, which we will
discuss in detail in Sec.~\ref{sec:impmodel}.
It can be measured, following Ref.~\onlinecite{CTQMC_Haule},
by their relative contribution to the trace: if $\Gamma$ is a multiplet of $H_\mathrm{dimer}$,
we define the CTQMC statistical weight of $\Gamma$ as
\def\CCC{{\cal C}_n(a_i,\sigma_i,\tau_i,\tau'_i)}
\begin{align}
&p_\Gamma^\text{QMC} \equiv
 \moy{
 \dfrac {
\left |
 \moy {\Gamma  \left | {\cal T} e^{-\beta H_\mathrm{dimer}} \CCC \right |\Gamma}
\right |
 }
 {
\displaystyle
\sum_{\Gamma}
\left |
 \moy {\Gamma  \left | {\cal T} e^{-\beta H_\mathrm{dimer}} \CCC  \right |\Gamma}
 \right |}}_\text{\!\!\smash{QMC}}
\label{def:p_gamma_QMC}
\\
&\CCC \equiv  \CC,
\end{align}
where $\moy{A}_\text{QMC}$ denotes the Monte Carlo averaging over the configurations
(labeled by $n$ and $\{a_i,\sigma_i,\tau_i,\tau'_i\}$).

In this simple two-impurity model in the paramagnetic phase, the symmetry between the two sites
allows to factorize the determinants:~\cite{CTQMC_Werner,CTQMC_Haule} $\Delta$ is indeed diagonal in the $\pm$ basis.
We also note that the Monte Carlo sign is one in this problem.
Due to the efficiency of this algorithm, we can routinely do a few millions Monte Carlo
sweeps and obtain high-quality data in imaginary time.
We then perform the continuation to the real axis using a simple Pad{\'e} method~\cite{vidberg_pade}
(see also Sec.~\ref{3-SpectralFnt} and Appendix~\ref{app:pade}).

\subsection{Rotationally-invariant slave bosons}
\label{sec:risb}
%

\begingroup
\begin{table*}[t]
\caption{Eigenstates of the dimer. The quantum numbers for charge, spin, and parity are given.
The last column shows the slave bosons for the description of the eigenstates in
the RISB formalism. States $10$ and $11$ have all their quantum numbers equal and form a $2\times 2$ block.
The ground state is number 10 which is the antiferromagnetic singlet
of even parity and has an energy
$E_{10}=2\varepsilon_0 + \tfrac{1}{2}(U-{\sqrt{16\,\bar t^2 + U^2}})\simeq 2\varepsilon_0-4\bar t^2/U$.
Here $\ell_\pm=U\pm\sqrt{16\bar t^2 +U^2}/4\bar t$ and $\mathcal{N}_\pm =2+2\ell_\pm^2$.
\label{table-states}}
\begin{ruledtabular}
\begin{tabular}{ccccccccc}
No.& Label (cf. Fig.~\ref{fg:weights}) & Eigenstate        &$n_\uparrow$&$n_\downarrow$&Parity&S&Boson\\
1  & E & $|0,0\rangle$ & 0        & 0          & +    &0&$\phi_{1,1}$\\
2  & 1+\, $(S_z=+1/2)$ & $\tfrac{1}{\sqrt{2}}(|0,\uparrow\rangle + |\uparrow,0\rangle)$ & 1        & 0          & +    &1/2&$\phi_{2,2}$\\
3  & & $\tfrac{1}{\sqrt{2}}(|0,\uparrow\rangle - |\uparrow,0\rangle)$ & 1        & 0          & -    &1/2&$\phi_{3,3}$\\
4  & 1+\, $(S_z=-1/2)$ & $\tfrac{1}{\sqrt{2}}(|0,\downarrow\rangle + |\downarrow,0\rangle)$ & 0        & 1          & +    &1/2&$\phi_{4,4}$\\
5  & & $\tfrac{1}{\sqrt{2}}(|0,\downarrow\rangle - |\downarrow,0\rangle)$ & 0        & 1          & -    &1/2&$\phi_{5,5}$\\
6  & T\, $(S_z=+1)$ & $|\uparrow,\uparrow\rangle$ & 2        & 0          & -    &1&$\phi_{6,6}$\\
7  & T\, $(S_z=-1)$ & $|\downarrow,\downarrow\rangle$ & 0        & 2          & -    &1&$\phi_{7,7}$\\
8  & T\, $(S_z=0)$ & $\tfrac{1}{\sqrt{2}}(|\uparrow,\downarrow\rangle+|\uparrow,\downarrow\rangle)$ & 1        & 1          & -    &1&$\phi_{8,8}$\\
9  & & $\tfrac{1}{\sqrt{2}}(|0,\uparrow\downarrow\rangle-|\uparrow\downarrow,0\rangle)$ & 1        & 1          & -    &0&$\phi_{9,9}$\\
10  & S & $\tfrac{1}{\sqrt{\mathcal{N}_-}}(-|\uparrow,\downarrow\rangle+\ell_-|\uparrow\downarrow,0\rangle+\ell_-|0,\uparrow\downarrow\rangle+|\downarrow,\uparrow\rangle)$ &1&1&+&0&$\phi_{10,10}$; $\phi_{10,11}$\\
11  & & $\tfrac{1}{\sqrt{\mathcal{N}_+}}(-|\uparrow,\downarrow\rangle+\ell_+|\uparrow\downarrow,0\rangle+\ell_+|0,\uparrow\downarrow\rangle+|\downarrow,\uparrow\rangle)$ &1&1&+&0&$\phi_{11,10}$; $\phi_{11,11}$\\
12  & & $\tfrac{1}{\sqrt{2}}(|\uparrow\downarrow,\uparrow\rangle + |\uparrow,\uparrow\downarrow\rangle)$ & 2        & 1          & +    &1/2&$\phi_{12,12}$\\
13  & & $\tfrac{1}{\sqrt{2}}(|\uparrow\downarrow,\uparrow\rangle - |\uparrow,\uparrow\downarrow\rangle)$ & 2        & 1          & -    &1/2&$\phi_{13,13}$\\
14  & & $\tfrac{1}{\sqrt{2}}(|\uparrow\downarrow,\downarrow\rangle + |\downarrow,\uparrow\downarrow\rangle)$ & 1        & 2          & +    &1/2&$\phi_{14,14}$\\
15  & & $\tfrac{1}{\sqrt{2}}(|\uparrow\downarrow,\downarrow\rangle - |\downarrow,\uparrow\downarrow\rangle)$ & 1        & 2          & -    &1/2&$\phi_{15,15}$\\
16  & & $|\uparrow\downarrow,\uparrow\downarrow\rangle$ & 2        & 2          & +    &0&$\phi_{16,16}$\\
\end{tabular}
\end{ruledtabular}
\end{table*}
\endgroup

%

Slave-boson (SB) methods (see e.g. Ref.~\onlinecite{Barnes1976,coleman_prb_1984,Kotliar1986,Lechermann2007a})
provide a simplified
description of the low-energy excitations in strongly-correlated electron systems.
The general idea is to enlarge the original Hilbert space to a set of states which involve
both (`slave') bosonic and (`quasiparticle') fermionic variables. The bosonic variables are introduced in such a way
that the interaction term becomes a simple quadratic form in terms of the slave bosons.
The physical Hilbert space is recovered by imposing a (quadratic) constraint relating
the slave bosons to the fermions.
Mean-field (approximate) solutions can be obtained by looking for saddle points at which the bosons
are condensed and at which the constraint is only satisfied on average.
At such a saddle point, the fermionic variables can be interpreted as the low-energy quasiparticles
of the original problem. A very simple form of the self-energy is obtained, containing
only a constant term and a term linear in frequency. This should be interpreted as a simplified
low-energy description of the system (i.e. as the first two terms in a low-frequency
expansion of the self-energy).

Because of the simplicity of this physical interpretation, slave-boson methods
are a very complementary tool to fully numerical algorithms such as the one
reviewed in the previous section. For this reason, we have also considered a SB
mean-field solution of the VB-DMFT equations, which should be viewed as a simplified
description of the low-energy physics. The full numerical solution (obtained with
CTQMC) has of course a rich frequency dependence but, as we shall see, the
SB approximation compares very well to the numerical results at low energy.

To be specific, we use a slave-boson mean-field as an approximate `impurity solver' of
the effective two-impurity problem coupled to self-consistent baths, within the
self-consistency iterative loop of VB-DMFT.
%
%
We use the recently introduced `rotationally-invariant' slave-boson
formalism~\cite{Lechermann2007a} (RISB), which generalizes the
original construction of Kotliar and Ruckenstein~\cite{Kotliar1986} to multi-orbital
systems in a way which respects all symmetries of the Hamiltonian
(see also Refs.~\onlinecite{li_spin-rotation-invariant_1989,fabrizio_gutzwiller_2007}).
To define the slave-boson variables, we consider the (`molecular')
eigenstates $|\Ga\rangle$
of the Hamiltonian~(\ref{def:HDimer}) describing the isolated 2-site cluster in the
absence of the baths. For definiteness, these $16$ states and their quantum numbers
are listed in Table~\ref{table-states}.
In the original formulation of Kotliar and Ruckenstein,~\cite{Kotliar1986}
a slave boson $\phi_\Ga$ is introduced for each molecular eigenstate.
However, this breaks rotational invariance in spin and orbital space, and
results into difficulties when saddle-point solutions are considered.
For example, nonequivalent saddle-point solutions would be found depending on whether
the Hamiltonian is expressed in the $(1,2)$-basis (corresponding to sites) or
in the $(+,-)$ basis corresponding to the even and odd orbital.
To avoid this problem, the RISB~\cite{Lechermann2007a} formalism introduces
a {\it matrix} of slave-boson amplitudes, $\{\phi_{\Gamma\Gamma^\prime}\}$.
The first index ($\Ga$) is associated with each molecular eigenstate in the
physical Hilbert space of the dimer. The second index ($\Ga'$) corresponds to
a state of the quasiparticle fermionic variables (with the same fermionic content
as the corresponding physical state).
Symmetry considerations allow for a drastic reduction of the number of
bosons to be considered in practice: Only the $\phi_{\Ga\Ga'}$ such that both
states have identical quantum numbers take non-zero values at the saddle point.
For the problem at hand, this leaves $18$ boson amplitudes in total, all scalars except for
a $2\times 2$ block in the two-particle sector with total spin $S=0$ and even parity
(Table~\ref{table-states}). Other bosons turn out to be zero at the
mean-field level.
%


To exclude the nonphysical states we impose the following set of constraints
%
%
\begin{eqnarray} \label{eq:constr0}
&&\sum_{\Ga\Ga^\prime}
\phi_{\Gamma\Gamma^\prime}^\dagger
\phi_{\Gamma\Gamma^\prime}\,=\,1\\ \label{eq:constr1}
&&\sum_{\Gamma\Gamma_1^\prime \Gamma_2^\prime}\phi_{\Gamma\Gamma_1^\prime}^\dagger
\phi_{\Gamma\Gamma_2^\prime}\langle \Gamma_2^\prime|f^\dagger_\al f_\be|\Gamma_1^\prime\rangle\,=\,
f^\dagger_\al f_\be\,\,\,,\,\,\,\forall\,\alpha,\beta,
\end{eqnarray}
where $f^\dagger_\al$ creates a quasiparticle in orbital $\alpha$.
For the dimer, we have
$\alpha=\{\pm,\sigma\}$ where $\pm$ designates the
even/odd orbital, $\sigma=\uparrow,\downarrow$ is the spin index and
$f_{\pm\sigma}=\frac{1}{\sqrt{2}}(f_{1\sigma}\pm f_{2\sigma})$.
The molecular eigenstates of the physical Hilbert space of the dimer have the
following representation,~\cite{Lechermann2007a} which satisfy the
constraints~(\ref{eq:constr0})
\begin{equation} \label{eq:state}
|\underline{\Gamma}\rangle\equiv \frac{1}{\sqrt{D_\Gamma}}\sideset{}{^\prime}\sum_{\Gamma^\prime}\phi^\dagger_{\Gamma\Gamma^\prime}|\text{vac}\rangle |\Gamma^\prime\rangle,
\end{equation}
where $D_\Gamma$ is a normalization factor and the primed sum is over states $|\Gamma^\prime\rangle$
which have the same quantum numbers as $|\Gamma\rangle$.
%
The physical electron operators can be expressed at saddle point as a
linear combination of the quasiparticle operators $f$ as
\begin{equation} \label{eq:physop}
d_\al^\dagger \to \underline{d}^\dagger_\al= R^*_{\al\be} f^\dagger_\be,
\end{equation}
where $R_{\al\be}$ is given by
\begin{equation} \label{eq:R}
R_{\alpha \beta}=\sum_{\Gamma_1\Gamma_2\Gamma_1^\prime\Gamma_2^\prime}\sum_\gamma
M_{\gamma\beta}\,
\langle \Gamma_1|d_\alpha|\Gamma_2\rangle
\langle \Gamma^\prime_1|f_\gamma|\Gamma^\prime_2 \rangle
\phi_{\Gamma_1\Gamma_1^\prime}^\dagger \phi_{\Gamma_2\Gamma_2^\prime}^{}.
\end{equation}
In this expression, $M_{\gamma\beta}$ is a matrix of normalization factors which insure that the
exact non-interacting solution is recovered in the saddle-point approximation when
$U=0$. Its explicit expression in terms of the slave-boson amplitudes can be
found in Ref.~\onlinecite{Lechermann2007a}.

Writing the dimer Hamiltonian as
\begin{equation}
\underline{H}_{\mathrm{dimer}}=
\sum_{\Gamma}E_\Gamma \sum_{\Ga^\prime}
\phi_{\Gamma\Gamma^\prime}^\dagger \phi_{\Gamma\Gamma^\prime}^{},
\end{equation}
we obtain the partition function as a functional integral
over coherent Bose and Fermi fields.
The constraints of Eq.~(\ref{eq:constr0}) are enforced including time-independent
Lagrange multipliers $\lambda_0$ and $\Lambda_{\alpha\beta}$.
The fermionic fields can be integrated out and the resulting bosonic action is treated
in the saddle-point approximation.
The free energy is finally obtained as
\begin{eqnarray} \label{eq:freenp}
\Omega &=&  -\frac{1}{\beta} \sum_{i\omega_n}\text{Tr}\ln[-\mathbf{G}_f^{-1}(i\omega_n)] -\lambda_0 +\\
&+& \sum_{\Gamma_1\Gamma_2\Gamma_1^\prime\Gamma_2^\prime}
\varphi_{\Gamma_1\Gamma_2^\prime}^*
\{
\delta_{\Gamma_1^\prime\Gamma_2^\prime}\delta_{\Gamma_1\Gamma_2} (\lambda_0 + E_{\Gamma_1})-\nonumber \\
&-&\delta_{\Ga_1\Ga_2}\sum_{\alpha\beta}\langle \Ga^\prime_1|
f_\al^\dagger \Lambda_{\al\be} f_\be
|\Gamma_2^\prime\rangle\}\,
\varphi_{\Gamma_2\Gamma_1^\prime}\nonumber.
\end{eqnarray}
In this expression, $\varphi_{\Gamma\Gamma^\prime}\equiv\langle\phi_{\Gamma\Gamma^\prime}\rangle$
are saddle-point (c-numbers) expectation values of the boson fields and
$G_f$ is the quasiparticle (auxiliary fermion) Green's function of the
impurity model, given by
\begin{equation}
\mathbf{G}_f^{-1}(\iomn)=
\iomn \mathbf{1} -\mathbf{\Lambda} - \mathbf{R}^\dagger\,
\left(
  \begin{array}{cc}
    \Delta_{+}(\iomn) & 0 \\
    0 & \Delta_{-}(\iomn) \\
  \end{array}
\right)
\mathbf{R},
\end{equation}
with $\Delta_\pm$ the hybridization function of the even (resp. odd) orbital.
The saddle-point approximation is obtained by extremalizing $\Omega$ over the boson
amplitudes $\varphi_{\Gamma\Gamma^\prime}$ and the Lagrange multipliers $\lambda_0,\Lambda_{\alpha\beta}$.

Within this approximation, the self-energy for the physical electron operators
consists simply in a constant term and a term linear in frequency, which are
orbital-dependent and read~\cite{Lechermann2007a}
\begin{equation}
\Sigma_d(\iomn)=\iomn\,(\mathbf{1}-[\mathbf{RR}^\dagger]^{-1})+[\mathbf{R}^\dagger]^{-1}\mathbf{\Lambda}\mathbf{R}^{-1}
-\varepsilon_0 \mathbf{1},
\end{equation}
so that the matrix of quasiparticle weights reads
\begin{equation}
\mathbf{Z}=\mathbf{R}\mathbf{R}^\dagger = \left(
                                            \begin{array}{cc}
                                              Z_{+} & 0 \\
                                              0 & Z_{-} \\
                                            \end{array}
                                          \right).
\end{equation}
Of course, as mentioned above, this form of the self-energy should be understood as a low-energy
approximation retaining only the constant terms (which renormalize the level position associated
with the even- and odd- orbitals) and the quasiparticle weights. Lifetime effects, as well as
higher order terms in frequency are neglected within the SB mean-field approximation.
Accordingly, the physical electron Green's function
\begin{equation}
\mathbf{G}_d = \mathbf{R}\,\mathbf{G}_f\,\mathbf{R}^\dagger,
\end{equation}
retains only the quasiparticle contribution to the spectral functions (note
in particular that these spectral functions do not satisfy the normalization
sum rule, and instead have spectral weights $Z_\pm$ corresponding to the quasiparticle
contributions).

%
From the expression of the intra-dimer energy at the mean-field level
\begin{equation}
\langle H_{\mathrm{dimer}}\rangle=\sum_{\Gamma} E_\Gamma \sum_{\Gamma^\prime}|\varphi_{\Gamma\Gamma^\prime}|^2,
\end{equation}
we see that the quantity
\begin{equation}\label{def:p_gamma_RISB}
p^{\rm{SB}}_\Ga \equiv \sum_{\Gamma^\prime} |\varphi_{\Gamma\Gamma^\prime}|^2
\end{equation}
can be interpreted as the statistical weight associated with the
contribution to the low-energy physics of the multiplet state $\Gamma$.
Because of the constraint~(\ref{eq:constr0}) on the slave bosons, these weights are normalized
according to $\sum_\Ga p^{\rm{SB}}_\Ga =1$.
As discussed later in this article, they can be directly compared to the
statistical weights computed within the CTQMC algorithm.

\section{Orbital-selective Mott transition in momentum space}
\label{sec:osmt}

\subsection{The different regimes of doping and the transition}
\label{sec:transition}

%
%
%
%
%
%
%
%
%

\begin{figure}[ht!]
  \includegraphics[width=8.5cm,clip=true]{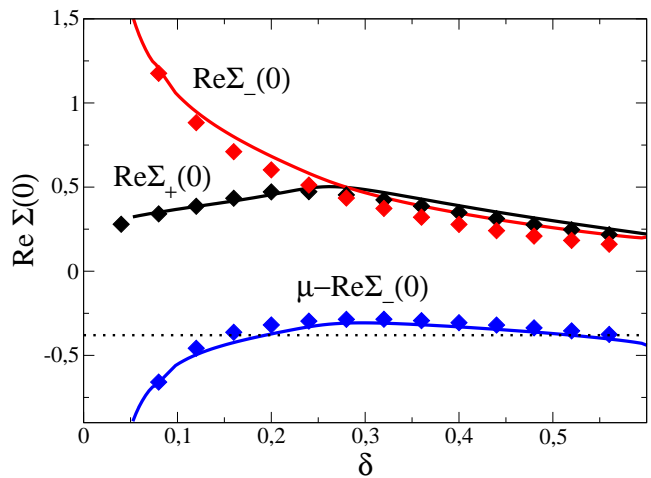}
  \caption{(Color online) Real part of the even and odd self-energies at $\omega=0$, extrapolated from CTQMC results
for $\beta= 200$, $U = 2.5$. The solid lines are the slave-boson (RISB) solution,
while the symbols are the CTQMC results.
The dotted line is the lower band edge of the $P_-$ patch
represented in Fig.~\ref{fg:partial_dos}.
}
  \label{fg:sigma_real}
\end{figure}
%
In this section, we describe the behavior of the system in the different regimes
of doping. First, we focus on a low-energy analysis which yields a very
simple description of the orbital-selective transition.
To this end, we first analyze the behavior of
the real part of the even- and odd-orbital self-energies, extrapolated to zero
frequency. This is shown in Fig.~\ref{fg:sigma_real} using both CTQMC and RISB.
Even though the full frequency dependence of the self-energy obtained
by CTQMC is highly non-trivial (see subsequent sections), its zero-frequency
limit is in remarkable agreement with the RISB solution, as is clear from
Fig.~\ref{fg:sigma_real}.

At large doping $\delta \gtrsim 20\%$, $\Sigma^\prime_{+}(0)$ and $\Sigma^\prime_{-}(0)$ are
very close to each other: Both orbitals behave in a similar way. In this
large doping regime, the system is a good metal with well-defined quasiparticles
everywhere on the Fermi surface. A single-site DMFT description is quite
accurate in this regime, since there is little orbital differentiation and hence
little momentum dependence.

As the doping level is further reduced ($\delta \lesssim 20\%$), the two orbitals
start to behave differently, signaling the onset of momentum differentiation
in the lattice model. The odd-orbital self-energy $\Sigma^\prime_{-}(0)$ increases
rapidly as the doping level is reduced, while $\Sigma^\prime_{+}(0)$ remains
much smaller and even decreases slightly with doping. At the critical doping level
$\delta\simeq 16\%$, the effective chemical potential of the odd orbital
$\mu - \Sigma_-^\prime(0)$ reaches the lower edge
$\epsilon_\mathrm{min} = -0.38$ of the non-interacting partial density of states
corresponding to the outer patch (Fig.~\ref{fg:partial_dos}), as signaled by the
dashed horizontal line in Fig.~\ref{fg:sigma_real}.
Retaining only the real part of the self-energy, this implies that the
pole equation corresponding to the outer patch
$\omega+\mu - \ek - \Sigma_{-}^\prime(\omega)=0$ no longer has solutions
for $\omega=0$, signaling the disappearance of low-energy quasiparticle
excitations from the outer patch.
%
%
Hence, for $\delta\lesssim 16\%$, we have a strongly
momentum-differentiated metal, with quasiparticles present only within the
central patch.
Only when the doping eventually reaches $\delta = 0$ do these
quasiparticles in turn disappear, corresponding to a Mott insulator.
Note that in the
above analysis we neglected the contribution from the imaginary part of the
self-energy $\Sigma_{-}^{\prime\prime}(0)$. Indeed, our data indicates
that $\Sigma_{-}^{\prime\prime}(0)$ vanishes as the temperature goes
to zero, so that the transition does exist in this limit.  At finite temperatures,
the imaginary part of the self-energy gives a small contribution to the
spectral density at the chemical potential, but still a rapid change of
behavior is expected around the transition.

\begin{figure}[ht!]
  \includegraphics[width=8.5cm,clip=true]{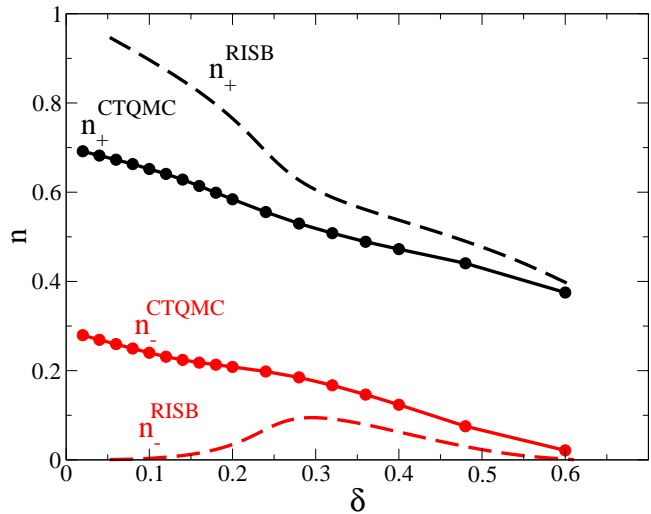}
  \caption{(Color online) Averages occupancies of the even
and odd orbitals, obtained with RISB method (dashed lines) and CTQMC (solid lines with circles).
$\beta= 200$, $U = 2.5$.
}
  \label{fg:occupations}
\end{figure}
%

Further insights into this transition can be obtained by analyzing the average occupancies in
each orbital $n_{+}$ and $n_{-}$ obtained by CTQMC and within the RISB
calculation, see Fig.~\ref{fg:occupations}. At large doping, the occupancies
obtained by both methods behave similarly and increase with decreasing doping.
As the doping gets closer to the critical value $\delta \simeq 16\%$, the RISB
solution displays a strong deviation where $n_{+}^{\rm{RISB}}$ increases rapidly and
$n_{-}^{\rm{RISB}}$ vanishes. Recalling that the slave-boson approximation only
accounts for the low-energy physics associated with quasiparticles,
this indicates that the odd orbital becomes empty \emph{at low energy} in the
low-doping phase, and that there
are no low-energy excitations left in the outer patch, as discussed
above. A change of behavior in $n_{\pm}$ at the transition is also
present in the CTQMC solution, but it should be kept in mind that these
quantities then include contributions from the higher-energy features
of the spectral function, and hence $n_{-}^{\rm{CTQMC}}$ is not expected to vanish
in the low-doping phase because the odd orbital does have spectral weight
at sufficiently negative energies in that phase as well.

\begin{figure}[ht!]
  \includegraphics[width=8.5cm,clip=true]{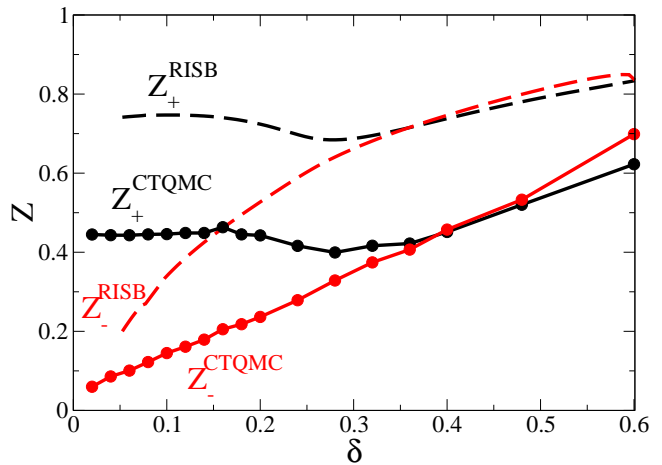}
  \caption{(Color online)
Quasiparticle residues of the even and odd orbitals obtained with RISB method (dashed lines)
and CTQMC (solid lines with circles).
$\beta= 200$, $U = 2.5$.
}
  \label{fg:residues}
\end{figure}
%
In our VB-DMFT approach, the strong differentiation in momentum space at low
doping manifests itself as an \emph{orbital-selective transition}: As one
approaches the Mott insulator, the odd orbital localizes at a finite doping level,
while the even one only does so at $\delta = 0$ when reaching the Mott insulator.
This is actually a crude description of the formation of
Fermi arcs. Indeed,
for $\delta < 16\%$, quasiparticles are only present in the inner patch, close
to the nodal region. Instead, in the antinodal region, the Fermi surface is
destroyed and the spectral function vanishes at the chemical potential. A more
precise description of the actual formation of the Fermi arcs requires to specify a
procedure for reconstructing the momentum dependence of the self-energy from this
two-orbital description: This is the topic of Sec.~\ref{sec:arcs}.

A marked difference of behavior between the two orbitals at low doping is
also found for the quasiparticle residues (see Fig.~\ref{fg:residues}) defined by
\begin{equation}
  Z_\pm = \Big( 1 - \frac{d\Sigma_{\pm}^{\prime}(\omega)}{d\omega}\Big|_{\omega \rightarrow 0} \Big)^{-1}.
\end{equation}
The CTQMC data and RISB approximation for $Z_\pm$ differ in absolute value but
they both display similar trends.
Again, at high doping $Z_{+}$ and $Z_{-}$are close to each other.
As the doping is reduced, $Z_{-}$ decreases (with roughly a linear dependence on doping)
while $Z_{+}$ remains essentially constant. Below the critical doping,
$Z_{-}$ cannot be interpreted as the spectral weight of a quasiparticle (the odd orbital is
localized), but it does indicates that the correlations continue to affect the
odd-orbital self-energy. Hence, correlations preferentially act on the antinodal
electrons. In contrast, correlations appear to have little influence on $Z_{+}$
below $\delta_c$, indicating that the nodal quasiparticles appear to be
``protected'' by the opening of the (pseudo-) gap in the antinodal regions.

\begin{figure}[ht!]
  \includegraphics[width=8.5cm,clip=true]{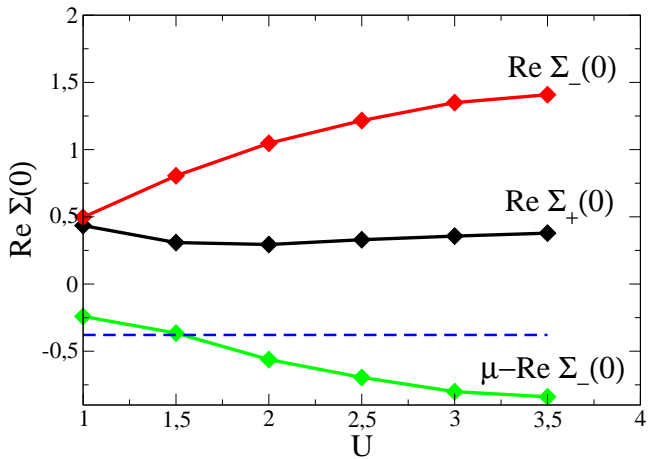}
  \caption{(Color online) Real parts of the self-energies at $\omega=0$, extrapolated from CTQMC results
at $\beta= 200$,  as a function of $U$ at a fixed doping $\delta = 8\%$.
The dotted line is the lower band edge of the $P_-$ patch
represented in Fig.~\ref{fg:partial_dos}.
}
  \label{fg:real_sigma_vs_U}
\end{figure}
%
The value of the critical doping $\delta_\mathrm{c}$ at which the transition
appears depends on the value of the interaction $U$. The larger $U$, the larger
$\delta_\mathrm{c}$. To illustrate the effect of $U$, we plot, in
Fig.~\ref{fg:real_sigma_vs_U}, the real parts of the self-energies extrapolated
to zero frequency for different values of $U$ at a fixed doping $\delta = 8\%$.
The difference between the even and the odd orbital increases with $U$. Above $U \simeq
1.5$, the renormalized chemical potential falls below the lower edge of the
partial DOS for the outer patch and the odd spectral function is vanishing at
the chemical potential. However, when $U < 1.5$, the odd orbital is metallic
again, showing clearly that the Coulomb interaction is at the origin of the
differentiation in momentum space.

\subsection{Spectral functions and the pseudogap at low doping}
\label{3-SpectralFnt}
%
%
%
%
%
%
%

\begin{figure}[ht!]
  \includegraphics[width=8.5cm,clip=true]{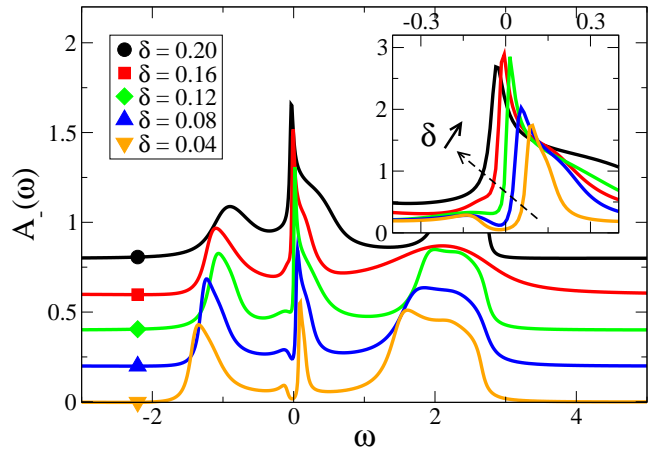}
  \caption{(Color online) Spectral function $A_{-}(\omega)$ for the odd orbital, obtained with Pad{\'e}
  approximants (see Appendix~\ref{app:pade}), for various dopings at $\beta =200$.
  A shift of 0.3 has been added between the curves for clarity.
  Inset: Zoom of the same curves at low frequencies (no shift added). }
  \label{fg:spectral_odd}
\end{figure}
%
%
In the previous section, we have shown that strong orbital differentiation sets in
at low-doping levels $\delta\lesssim 16\%$. In a simplified low-energy
description, the effective chemical potential for the odd orbital
is pushed below the lower band edge. This
corresponds to the vanishing of the low-energy spectral weight
of the odd orbital, and signals the disappearance of
low-energy quasiparticles in the antinodal regions. In this section, we
go beyond this simple low-energy analysis and study the full
frequency dependence of the spectral functions of both the even and odd
orbitals. One of the main outcomes of this study, as we shall see, is that
the odd orbital does not have zero spectral weight in a finite frequency
range around $\omega=0$, but rather develops a {\it pseudogap}.

The computation of real-frequency spectral functions is made possible
by the very high quality of the CTQMC results on the Matsubara axis, allowing for
reliable analytical continuations to the real axis at low and intermediate
energy, using simple Pad{\'e} approximants~\cite{vidberg_pade} (see Appendix~\ref{app:pade}).
This is a definite advantage of our simplified two-orbital approach, in which
the statistical noise of Monte Carlo data can be reduced down to very small
values at a reasonable computational cost.
In Fig.~\ref{fg:spectral_odd}, we plot the spectral function $A_{-}(\omega)$ of
the odd orbital at a fixed interaction $U=2.5$.
At high energies, the spectra display the expected lower and upper Hubbard
bands, and from now we focus on the lower energy range. In this range,
the spectra display a central peak. At high doping, this peak is centered at the
Fermi level $\omega=0$. As the doping level is reduced, this peak
shifts toward positive energies.
At the critical doping $\delta_c
\simeq 16\%$, the chemical potential is at the lower edge of the peak, in
agreement with the low-energy analysis discussed above.

Correspondingly, the spectral weight at $\omega=0$ is strongly
suppressed as the doping is reduced from $\delta_c
\simeq 16\%$. A pseudogap is formed at low energy,
as clear from the inset of Fig.~\ref{fg:spectral_odd}, which
deepens as the doping level is reduced.
There is no coherent spectral weight at the chemical potential. The finite
spectral weight at $\omega=0$ is due to thermal excitations.
In contrast, the finite spectral weight at small but non-zero frequency
survives as temperature is reduced, corresponding to a
pseudogap rather than a true gap in $A_{-}(\omega)$.

\begin{figure}[ht!]
  \includegraphics[width=8.5cm,clip=true]{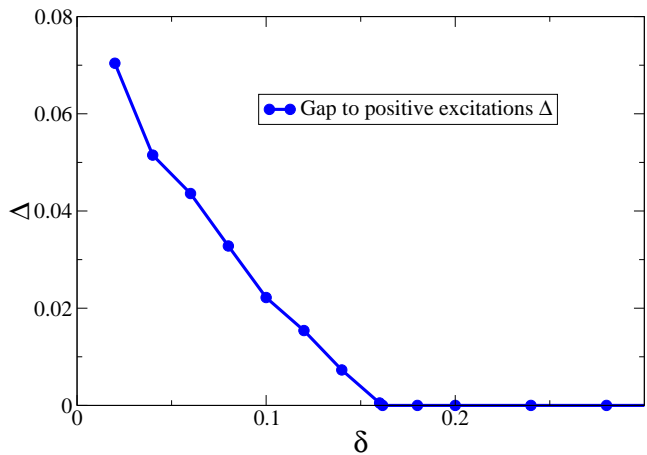}
  \caption{(Color online) Gap to {\it positive} coherent excitations in the odd orbital,
  obtained from Eq.~(\ref{EqForDeltaSup}).  }
  \label{fg:gap_odd}
\end{figure}
%
The prominent peak at low energies in $A_{-}(\omega)$ is associated with the first
coherent excitations at positive energies. By neglecting the effect of the
imaginary part of the self-energy, it is possible to precisely identify the position of
this peak as the scale $\Delta$
where the first positive-energy poles
appear in the expression of the odd-orbital Green's function
\begin{equation}
  G_{-}(\omega) =
    \sum_{\vk \in \mathcal{P}_{-}}
    \frac{1}{\omega + \mu - \ek - \Sigma_{-}^\prime(\omega)}.
\end{equation}
Hence, $\Delta$ is the solution of
\begin{equation}\label{EqForDeltaSup}
    \Delta + \mu - \epsilon_\mathrm{min} - \Sigma_{-}^\prime(\Delta) = 0,
\end{equation}
where $\epsilon_\mathrm{min}$ is the lower-band edge of the outer-patch partial
DOS. The solution of this equation is shown in Fig.~\ref{fg:gap_odd}.  The gap
$\Delta$ opens below $\delta_c$ and provides a characteristic energy scale for
the position of the peak, see inset in Fig.~\ref{fg:spectral_odd}. Note that
this energy scale is much smaller than the deviation of the renormalized
chemical potential $\mu - \Sigma_{-}^\prime(0)$ from the lower outer-patch
band edge because of the non-trivial frequency behavior of the self-energy.
Furthermore, the magnitude of $\Delta$ as obtained from Fig.~\ref{fg:gap_odd}
is in the range of tens of m$eV$'s consistent with the typical magnitude of the
pseudogap in cuprates.

\begin{figure}[ht!]
  \includegraphics[width=8.5cm,clip=true]{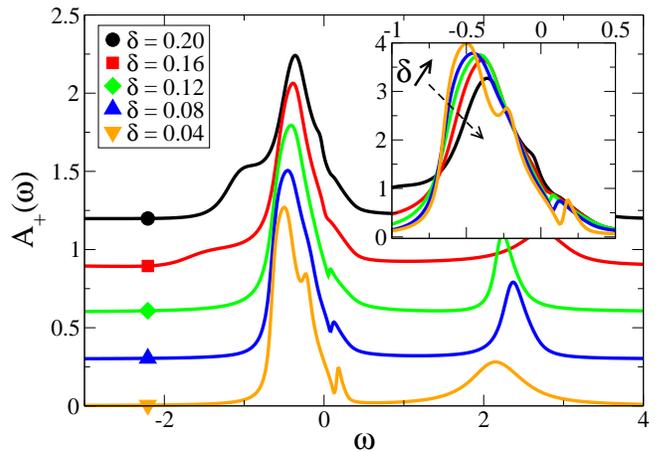}
  \caption{(Color online) Spectral function $A_{+}(\omega)$ for the even orbital, obtained with Pad{\'e}
approximants (see Appendix~\ref{app:pade}), for various dopings at $\beta =200$.
A shift of 0.2 has been added between each curves for clarity.
Inset: Zoom of the same curves at low frequencies (no shift added).  }
  \label{fg:spectral_even}
\end{figure}
%
In Fig.~\ref{fg:spectral_even}, we display the spectral function
$A_{+}(\omega)$ of the even orbital for different doping levels.
The dependence of $A_{+}(\omega)$ on doping is rather weak.
The main feature of the non-interacting density of states corresponding to
the central patch (Fig.~\ref{fg:partial_dos}) is recovered on these spectra,
namely a broad peak centered at negative energy with a tail leaking above the
Fermi level. The absence of a visible lower Hubbard band, as well as the relatively
small spectral weight of the upper Hubbard band (at the same position $\omega\simeq 2$
as in $A_{-}(\omega)$), indicate that correlations
have a much weaker effect on the even orbital (central patch, nodal regions) than on the
odd (antinodal) one, as already anticipated in the previous section.
The spectral function $A_{+}(\omega)$ is quite asymmetric, with more
hole-like excitations than particle-like excitations (in line with the
fact that the central patch corresponds mainly to filled states).
At low doping, a small dip appears close to the chemical potential. The
position of this dip is close to that of the prominent peak in the odd-orbital spectral
function.

\subsection{Comparison with tunneling experiments}

\begin{figure}[ht!]
  \includegraphics[width=8.5cm,clip=true]{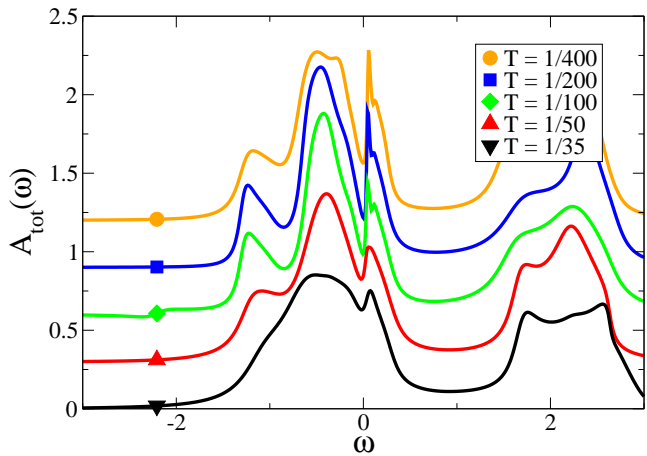}
  \caption{(Color online) Total spectral function $A_\mathrm{tot}(\omega)$
for various temperature at $\delta = 0.08$.
A shift of 0.3 has been added between each curves for clarity.  }
  \label{fg:spectral_total}
\end{figure}
%
A direct comparison can be made between our VB-DMFT cluster calculations and
tunneling experiments in the normal state of cuprate superconductors. Indeed,
tunneling directly probes the momentum-integrated spectral density, and hence
the comparison is free of the possible ambiguities associated with momentum-space
reconstruction which influence the comparison of cluster calculations to
momentum-resolved spectroscopies (as discussed in more detail in Sec.~\ref{sec:arcs}).
The tunneling conductance $dI/dV$ as a function of the voltage $V$ is given
by~\cite{ReviewSTMFisherBerthod}
%
%
\begin{equation}
\frac{dI}{dV} \propto \,\int_{-\infty}^{+\infty} d\omega \left[-f^\prime(\omega -eV)\right]
A_\mathrm{tot}(\omega) \label{eq:stm}.
\end{equation}
In this expression, tunneling between a normal metal (with a featureless density
of states) and the correlated sample is considered, $f^\prime$ designates the
derivative of the Fermi function, $e>0$ is the absolute value of the electron charge
and $A_\mathrm{tot}(\omega)$ is the local (momentum-integrated) spectral function.
The energy dependence of tunneling matrix elements has been neglected, and the correlated
sample is considered to be homogeneous.

In Fig.~\ref{fg:spectral_total}, we display $A_\mathrm{tot}(\omega) = A_{+}(\omega) + A_{-}(\omega)$
for different temperatures $T = 1/\beta$ at a fixed doping $\delta = 8\%$.
In this local spectral function, we recognize the features discussed above, namely
the broad band at negative energy originating from $A_{+}(\omega)$, and the sharp peak
at a small positive energy found in $A_{-}(\omega)$, separated by the pseudogap
at low energy.
%

\begin{figure*}[!ht]
  \includegraphics[width=8.5cm,clip=true]{stm_vs_T}
  \includegraphics[width=5cm,clip=true]{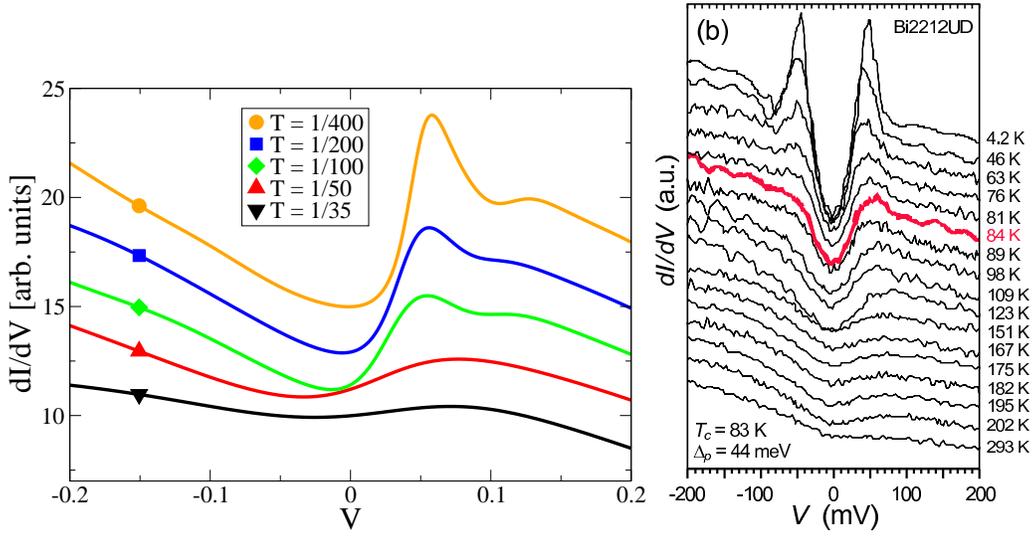}
  \caption{(Color online) Left panel: STM curve for various temperature at $\delta = 0.08$, in arbitrary units.
A shift has been added between each curves for clarity.
Right panel: DOS measured by STM experiments on Bi2212
with $T_c=83K$. Figure reprinted with permission from Ref.~\onlinecite{ReviewSTMFisherBerthod}
(Fig.~22b). Copyright 2007 by the American Physical Society. }
  \label{fg:stm}
\end{figure*}
%
In Fig.~\ref{fg:stm}, we display the voltage dependence of the tunneling conductance
obtained using the spectral function $A_\mathrm{tot}(\omega)$ calculated within
VB-DMFT. The results are displayed at a fixed, low-doping level $\delta = 8\%$ in
the pseudogap regime, for different temperatures.
For comparison, we also display the experimental data of Renner \emph{et al.}~\cite{renner_stm,ReviewSTMFisherBerthod}
for underdoped Bi2212. When comparing the two set of curves, attention should be paid to the
fact that our calculation applies at this stage only to the normal state $T>T_c$.
Our calculation compares quite favorably to the experimental data, in several respects.
At low temperature, both the theoretical and experimental conductance displays
i) a dip at low voltage corresponding to the pseudogap ii) a peak at a small positive
voltage (corresponding to empty hole-like states) and iii) an overall particle-hole
asymmetry $dI/dV<0$ at negative voltage as well as at positive voltage above the peak,
as indeed expected in a doped Mott insulator. Furthermore, we observe that the temperature
dependence reveals the gradual buildup of the positive-voltage coherence peak as temperature
is lowered, as well as the gradual opening of the pseudogap at low voltage.
One aspect of our theoretical results which departs from the
experiments is the detailed shape of the conductance at negative voltage: In experiments
a more pronounced dip is visible, while our results rather display a gradual, linear-like
decrease.

Our results have direct implications for the interpretation of tunneling experiments, and
also suggest some further experiments to test these predictions. First, the coherence
peak at small positive voltage must be associated, according to our theory, mainly with
low-energy empty states in the antinodal regions. Second, the position of this peak is predicted to have
a definite doping dependence, tracking $\Delta$ in Fig.~\ref{fg:gap_odd} and hence
moving to higher energy as the doping level is reduced from `optimal' doping.

\subsection{Frequency-dependence of the self-energy and the inelastic scattering rates}

\begin{figure*}[ht!]
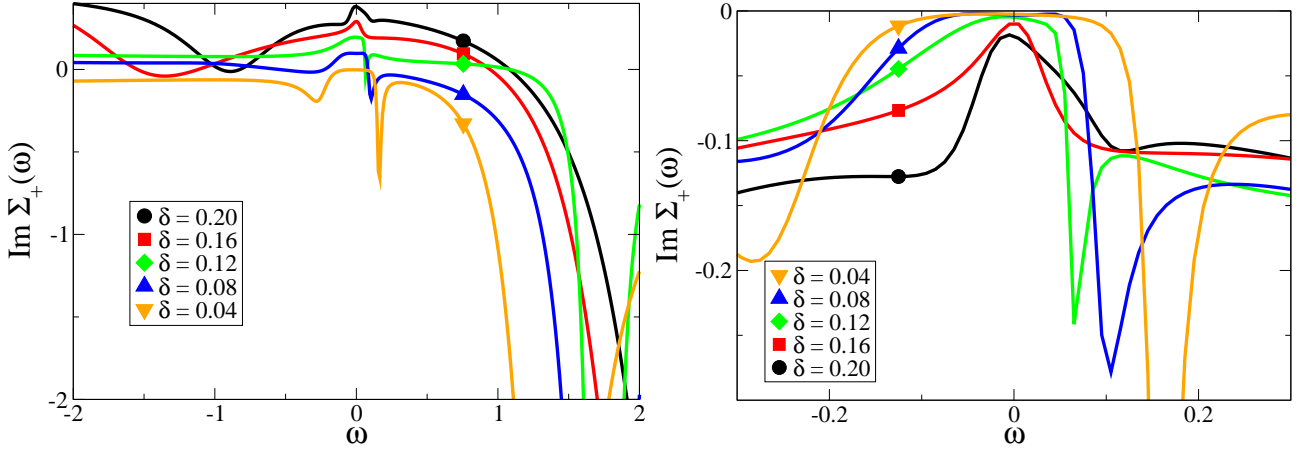

  \includegraphics[width=8.5cm,clip=true]{self_imag_even}
  \includegraphics[width=8.5cm,clip=true]{self_imag_even_zoom}
  \caption{(Color online) Left panel: Imaginary part of the self-energy $\Sigma_{+}(\omega)$ for the even orbital.
  A shift of 0.1 has been added between the curves for clarity. Right panel: Zoom over the low-frequency region
  (no shift added).}
  \label{fg:self_imag_even}
\end{figure*}
\begin{figure*}[ht!]
  \includegraphics[width=8.5cm,clip=true]{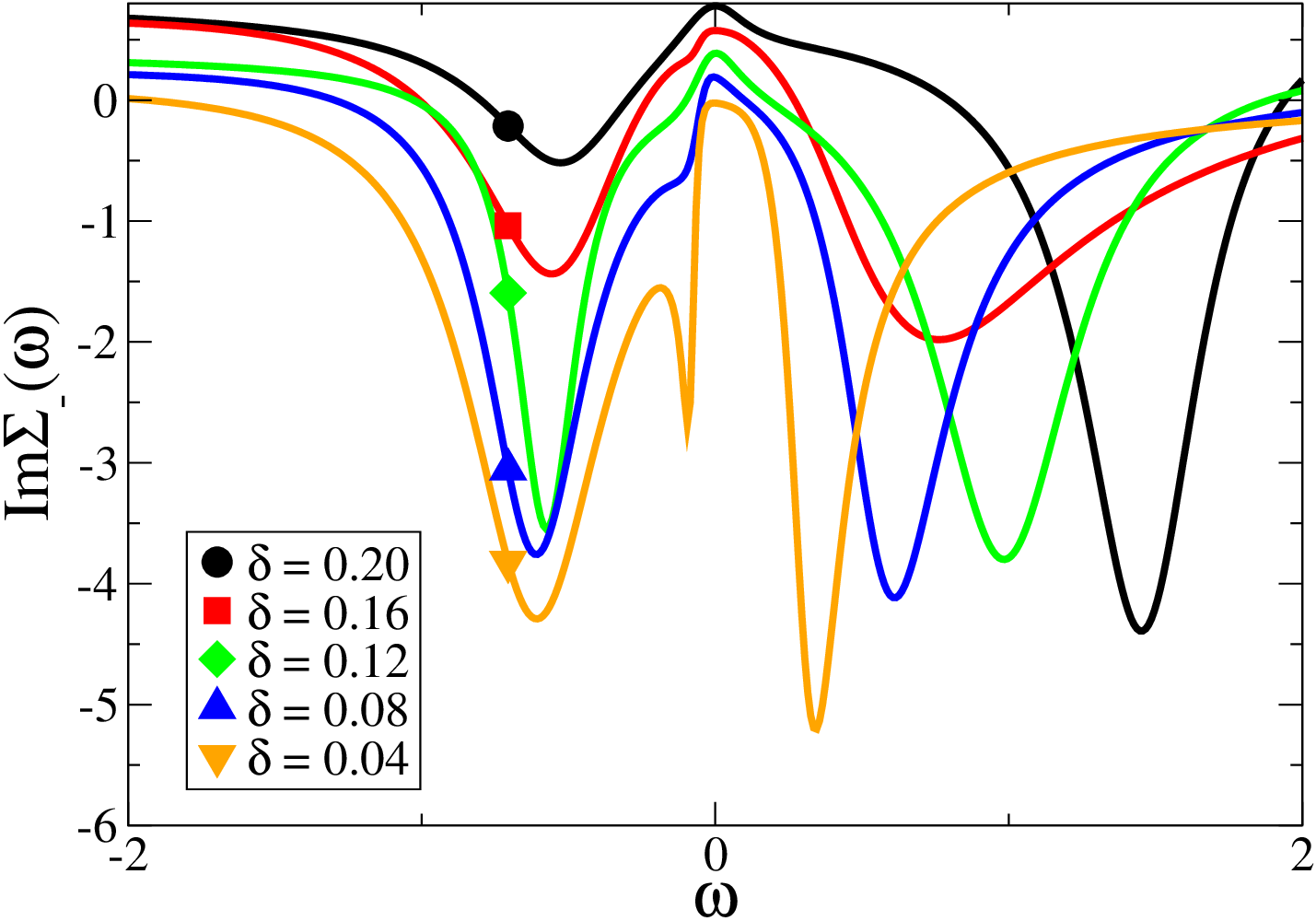}
  \includegraphics[width=8.5cm,clip=true]{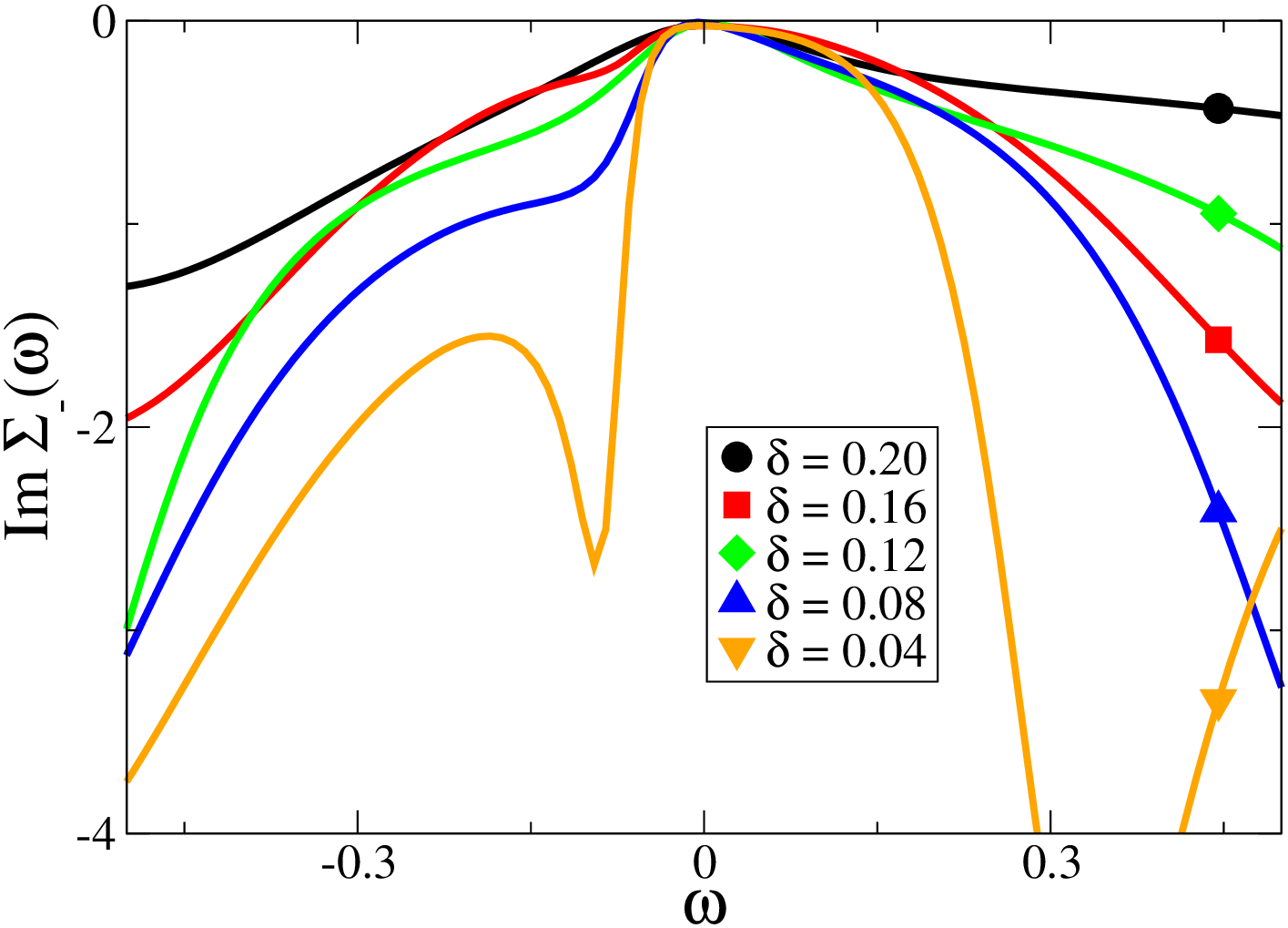}
  \caption{(Color online) Left panel: Imaginary part of the self-energy $\Sigma_{-}(\omega)$ for the odd orbital.
  A shift of 0.2 has been added between the curves for clarity. Right panel: Zoom over the low-frequency region
  (no shift added).}
  \label{fg:self_imag_odd}
\end{figure*}

Here, we discuss the frequency dependence of the imaginary part of the
self-energies $\Sigma_{\pm}^{''}(\omega)\equiv\rm{Im}\Sigma_{\pm}(\omega+i0^+)$ and
its physical implications for the inelastic scattering rates of the nodal
and antinodal quasiparticles in the different regimes of doping.
These quantities are displayed on Figs.~\ref{fg:self_imag_even},~\ref{fg:self_imag_odd}.
Let us recall that these quantities are directly related to the quasiparticle
lifetimes, which is given by the inverse of $Z_\pm \Sigma_{\pm}^{''}(\omega)$.

Again, we observe that at large doping, these quantities have rather similar behavior.
An approximately quadratic frequency dependence is found at low energy, corresponding
to a Fermi liquid behavior of both orbitals, and the self-energies display high-energy
peaks corresponding to the structures in the spectral functions described above. Overall,
the self-energies at large doping are quite similar to those found in the single-site DMFT
description of a correlated Fermi liquid.

The situation becomes radically different as the doping level is reduced.
The first observation is that the overall scale for
$\Sigma_{+}^{''}(\omega)$ and for $\Sigma_{-}^{''}(\omega)$ then becomes very
different. Clearly, away from the very low-energy region, $\Sigma_{-}^{''}(\omega)$ becomes
much larger than $\Sigma_{+}^{''}(\omega)$, indicating again a stronger effect of correlations
on the antinodal regions (odd orbital) than on the nodal ones (even orbital), and a
much larger degree of coherence of the nodal quasiparticles.

Focusing on the even orbital (nodes) at low frequency, we observe that
$\Sigma_{+}^{''}(\omega=0)$ displays a marked {\it decrease} as the doping level is
reduced from the characteristic doping $\delta_c\simeq 16\%$ at which orbital
differentiation sets in and the pseudogap opens. Physically, this means
that the opening of the pseudogap leads to a protection of the nodal quasiparticles
by increasing their inelastic lifetime at low energy. Indeed,
$\Sigma_{+}^{''}(\omega)$ displays a quite remarkable shape at low doping, with a rather
large interval of frequency around $\omega=0$ in which it is very small and flat,
indicating almost free nodal quasiparticles at low doping.

This is in marked contrast to the behavior of the odd (antinodal) orbital. In this case,
our real-frequency data lack the precision required to assess precisely the doping
dependence of the very low-frequency rate $\Sigma_{-}^{''}(\omega=0)$. However, as soon as
one focuses on a small but finite frequency (which indeed is relevant to the
lifetime of antinodal quasiparticles at the edge of the pseudogap), it is apparent from
Fig.~\ref{fg:self_imag_odd} that $\Sigma_{-}^{''}(\omega)$ rapidly {\it increases} as the
doping is reduced from $\delta_c$. This corresponds to increasingly incoherent antinodal
quasiparticles at low doping level.

Making contact with experiments, these observations appear to be in good qualitative
agreement with the fact that the in-plane resistivity of cuprate superconductors is
reduced when the pseudogap opens and that nodal quasiparticles survive at low
doping while the antinodal ones loose their coherence.

%
%
%
%
%
%

\subsection{Other patches}
\label{sec:otherpatches}


%
%

So far we have presented results for a particular patching scheme of
the BZ. 
The motivation behind this choice is based on the known phenomenology of the
cuprates: The central patch is shaped in such a way as to contain the
nodal point while the outer patch is modeled to contain
the incoherent antinodal region. Phenomenological
patch models using related patches have been used
to parametrize the transport properties of
the cuprate superconductors.~\cite{perali_epjb_2001}
%
%
As we saw, reducing the doping induces a transition in which the outer patch becomes insulating.
Here we address the stability of this picture with respect to the deformation of the patches.
We consider patches that do not break the lattice point symmetry and have equal
volume not to incur into problems with the definition of the cluster
Hamiltonian (for details on the formalism see Appendix~\ref{Appendix-dca}).

\begin{figure}
  \includegraphics[width=0.45\textwidth,clip=true]{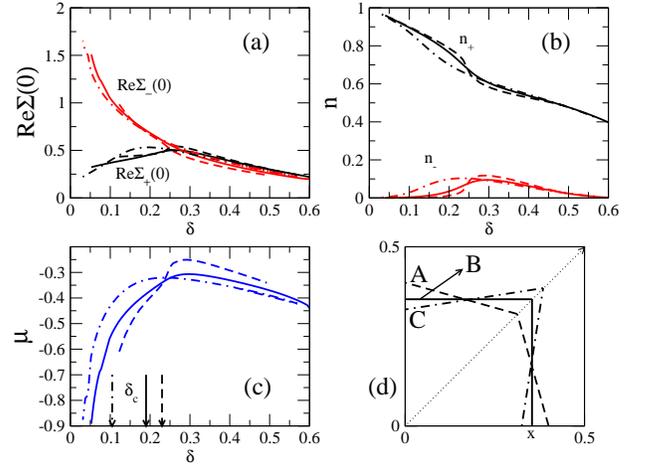}
  \caption{Doping dependence of (a) the real part of $\Sigma_\pm(0)$, (b) the even and odd orbital occupations, and
(c) $\mu-\Sigma_-(0)$. The arrows in (c) indicate the critical doping where the odd band becomes
insulating. Results for different patching schemes as shown in (d) using the RISB method. }
  \label{fig:patches_comparison}
\end{figure}

The main difference between the patching schemes that we consider (see Fig.~\ref{fig:patches_comparison}d)
is the relative weight given to the nodal and antinodal regions.
Compared to the reference patching used in the rest of the paper (patching B, also in Fig.~\ref{fg:patches}),
in patching A the central patch includes the node and also
large part of the antinode. At the opposite, in patching C the outer patch
has a larger contribution of the node.

In Fig.~\ref{fig:patches_comparison}a-c we present results for the three patching schemes of Fig.~\ref{fig:patches_comparison}d.
The results for the self-energy and the occupations are qualitatively very similar for the different patching schemes.
%
Most importantly, the distinctive feature of a selective insulating transition as the doping is decreased remains in all the cases.
An analysis of $\mu - \mathrm{Re} \Sigma_-(0)$ allows us to calculate the critical doping as indicated by the arrows in Fig.~\ref{fig:patches_comparison}c.

We notice how, by increasing the portion of the nodal region contained in the {\em outer}
patch, the critical doping shifts systematically to lower values. This trend is
consistent with a picture in which the quasiparticles are restricted to a
fraction (an arc) of the Fermi surface located near the node. By decreasing the
doping the quasiparticles disappear at the sides of the arc inducing the
selective transition in the outer orbital. Despite the similarity of the trend
with the experimental behavior in cuprates, the sensitivity of the method is
not sufficient to obtain a quantitative prediction for the size of the arc
as a function of doping.

We also investigated the real-space cluster method CDMFT
(cellular-DMFT).~\cite{kotliar_review_rmp_2006,maier_cluster_rmp_2005} In this method,
the notion of patches in the Brillouin zone cannot be introduced because the
starting point is a real-space cluster that breaks translational symmetry.
Nevertheless the qualitative picture emerging from CDMFT is very similar to
the one found in DCA. In particular, we find also in CDMFT a critical doping
below which the system is in a selective Mott-insulating state.

\section{Physics of the effective two-impurity model and the role of the self-consistency}
\label{sec:impmodel}

In this section, we relate and contrast the orbital-selective transition described above to
the physics of the crossover between a Kondo-dominated and a singlet-dominated regime
in the two-impurity Anderson model.

\subsection{Singlet dominance at low doping}

%
%

\begin{figure}[ht!]
  \includegraphics[width=8.5cm,clip=true]{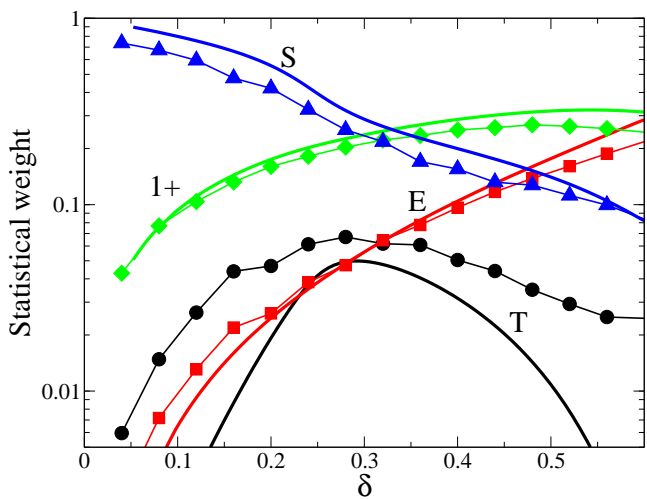}
  \caption{(Color online) Statistical weights of the various dimer cluster eigenstates (labeled
  as in Table~\ref{table-states}). $S$
  is the intra-dimer singlet, $1+$ the (spin-degenerate) state with one electron in the even orbital, $E$ the
  empty state and $T$ the intra-dimer triplet. $\beta = 200$.}
  \label{fg:weights}
\end{figure}
%
A way to obtain a more transparent physical picture of the phases encountered as a
function of doping is to study the contribution of the different cluster eigenstates
to the density matrix.
We plot in Fig.~\ref{fg:weights}
the statistical contribution of several cluster eigenstates $|\Gamma\rangle$
(see Sec.~\ref{sec:ctqmc}) from CTQMC ($p_\Gamma^\text{QMC}$ defined
in Eq.~\ref{def:p_gamma_QMC}) and the RISB method ($p_\Gamma^\text{SB}$ defined
in Eq.~\ref{def:p_gamma_RISB}).


The agreement between CTQMC and RISB is
very good, and even quantitative for the two states with highest weights.
At large doping, the empty state and the two spin-degenerate states with one
electron in the even orbital dominate, as expected.  As doping decreases, these
states lose weight and the intra-dimer singlet prevails, reflecting the strong
tendency to valence-bond formation.
Therefore, the orbital (momentum)
differentiation at low doping is governed by intra-dimer singlet formation.
This situation is strongly reminiscent of the behavior of the two-impurity
Anderson model (2IAM).~\cite{deleo2004,Zhu2006}  In the following subsection we will compare the results
for the 2IAM and VB-DMFT to find the extent of this similarity.

\subsection{Role of the self-consistency: from a RKKY/Kondo crossover to a transition}

%
%
%
%


The 2IAM has been thoroughly studied by many authors
using a variety of methods.~\cite{jayaprakash_two-impurity_1981,jones_mean-field_1989,affleck_conformal-field-theory_1995,deleo2004,cornaglia_strongly_2005}
Here we will focus on a simplified version of the 2IAM where
the impurities are coupled directly through a hopping term $\bar{t}$.
In standard notation the Hamiltonian is given by
\begin{eqnarray*}
H_{2IAM} &=& U(n_{1\uparrow}n_{1\downarrow}+n_{2\uparrow}n_{2\downarrow}) + \varepsilon_0\sum_{\sigma\alpha}n_{\alpha\sigma}\\
&-&\bar{t} \sum_\sigma (d^\dagger_{1\sigma}d_{2\sigma}^{} +H.c.) -  V\sum_{\vk\sigma} (d^\dagger_{1\sigma} c_{\vk 1 \sigma}^{}+H.c.)\\
&-&  V\sum_{\vk \sigma} (d^\dagger_{2\sigma} c_{\vk 2 \sigma}^{}+H.c.) + \sum_{\vk \alpha \sigma} \ek c^\dagger_{\vk \alpha\sigma} c_{\vk \alpha\sigma}^{},
\end{eqnarray*}
where $U$ is the local repulsion and $\varepsilon_0$ the level energy.
Note that each impurity is coupled to an
independent electronic bath and that there are no crossed baths that would couple to both impurities.
We choose the baths to have a semi-elliptic density of states of half-bandwidth $D=1$ and
the hybridization $V = 0.5D$.
This model corresponds to the impurity model that is solved in VB-DMFT,
with the important difference that the baths are kept fixed and $\Delta_{12} = 0$.

\begin{figure}
  \includegraphics[width=0.45\textwidth,clip=true]{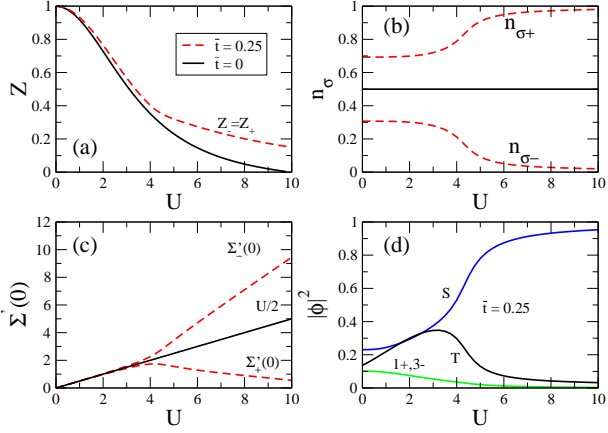}
  \caption{Kondo to RKKY crossover as a function of the coulomb repulsion $U$
at the impurities at half-filling.
(a) Quasiparticle weights.
(b) Even and odd orbital occupations.
(c) Real part of the self-energy.
(d) Boson amplitudes for the intra-dimer singlet (S), the intra-dimer triplet (T),
and the spin degenerate doublets with one electron and even symmetry (1+)
and three electrons with odd symmetry (3-).}
  \label{fig:Anderson2impU}
\end{figure}

For $\bar{t}=0$ the problem reduces to that of two independent single-impurity Anderson models.
The RISB method (as other slave-boson approaches) provides a description of the quasiparticles in the Kondo resonance.
The impurity spectral density has a single peak at the Fermi level whose width is of the order of the Kondo energy $ T_K$.
The real part of the self-energy in the electron-hole symmetric situation ($\varepsilon_0=-U/2$),
is simply $\mathrm{Re} \Sigma_\pm(0)=U/2$ while $Z$ decreases monotonously with increasing $U$ and exponentially for large $U$ (see Fig.~\ref{fig:Anderson2impU}).
When $\bar{t}$ is turned on, an antiferromagnetic coupling $I=4\bar{t}^2/U$  is generated between the impurities.
If $I \lesssim T_K$ (see Fig.~\ref{fig:Anderson2impU} for $U\lesssim 4$ ) both impurities are in the Kondo regime as in the $\bar{t}=0$ case
and the behavior of $\mathrm{Re} \Sigma_\pm(0)$ and Z reproduces that of uncoupled impurities.
In the opposite limit (large $U$ in Fig.~\ref{fig:Anderson2impU}) there is a large inter-impurity correlation which is signaled by a differentiation between $\mathrm{Re} \Sigma_+(0)$ and $\mathrm{Re} \Sigma_-(0)$ (i.e. the emergence of a $\mathrm{Re} \Sigma_{12}(0)$ term).

The behavior of the bosonic amplitudes with increasing interaction $U$ clearly shows the two regimes and the crossover region
(see Fig.~\ref{fig:Anderson2impU}d).
In the Kondo regime both the singlet and triplet multiplets have large amplitudes,
while in the so-called RKKY (Ruderman-Kittel-Kasuya-Yosida) regime the singlet dominates the physics and its
associated bosonic amplitude is close to one.
Since the level occupation is related to the bosonic amplitudes through the constraints of Eq.~(\ref{eq:constr0}),
their behavior follows.

In the RKKY regime, the even orbital (where the ground state singlet lives)
fills up while the odd orbital empties (see Fig.~\ref{fig:Anderson2impU}) as in
earlier studies of $SU(N)$ two-impurity  models in the large-$N$
limit.~\cite{jones_mean-field_1989}
The  RISB method brings a significant advantage over
previous methods since it generates at the mean-field level the RKKY
interaction which in previous treatments had to be introduced in an ad-hoc
fashion or by treating fluctuations in higher orders in
$1/N$.~\cite{houghton_prb_1988}

\begin{figure}
  \includegraphics[width=0.45\textwidth,clip=true]{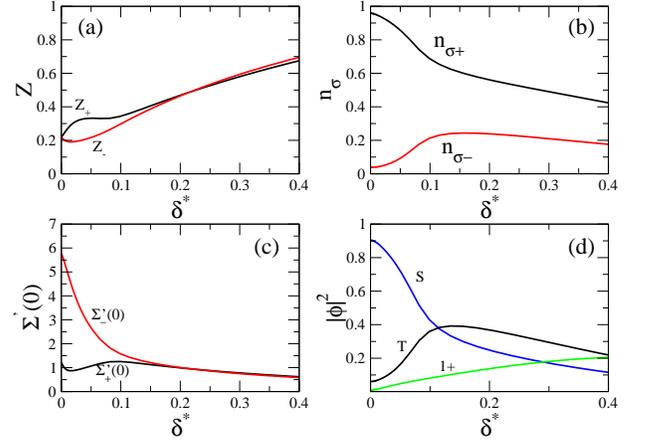}
  \caption{Kondo to RKKY crossover for the two impurity Anderson model
as described by RISB as a function of the impurity level doping $\delta^*=1-n$.
Parameters are $U=7D$ and $\bar{t}=0.25D$.
(a) Quasiparticle weights.
(b) Even and odd orbital occupations.
(c) Real part of the self-energy.
B
(d) Boson amplitudes for the intra-dimer singlet (S), the intra-dimer triplet (T),
and the spin degenerate doublet with one electron and even symmetry (1+).
}
  \label{fig:Anderson2impU7dop}
\end{figure}

The crossover can also be observed by changing the impurity occupation,
i.e. by shifting the local energy $\varepsilon_0$.
If Fig.~\ref{fig:Anderson2impU7dop}, we present  different physical quantities as a function of the
impurity doping level $\delta^*=1-\sum_{\alpha,\sigma}n_{\alpha\sigma}^{}$.
At zero doping the system is electron-hole symmetric and for the value of $U=7D$ in the figure
it is in the RKKY regime. This is clearly observed in the
boson amplitudes, the level occupations and the real part of the self-energy at zero frequency.
Increasing the doping increases the charge fluctuations in the impurity and this enhances the Kondo correlations.
At a doping level $\delta^*\sim 0.1$ there is a crossover to the Kondo regime where the inter-impurity correlation is small.
For large values of $\delta^*$ the impurities enter an empty orbital regime an the effect of correlations is small. 

The behavior observed as a function of doping for the different quantities of
the impurity model closely resembles that of VB-DMFT.
Note however that the 2IAM presents a crossover between the Kondo and RKKY regimes while in VB-DMFT there is an
orbital-selective Mott transition.
The origin of the transition can be traced back to the only difference between VB-DMFT and the 2IAM,
namely the presence in VB-DMFT of self-consistently determined baths.
Indeed the odd-orbital Green's function for the 2IAM is
\begin{equation}
   G_-(\omega) = \frac{1}{\omega + \mu - \Delta^\mathrm{hybr}_-(\omega) - \Sigma_-(\omega)},
\end{equation}
where $\Delta^\mathrm{hybr}_-(\omega) = 2 (V/D)^2 \int d \varepsilon \frac{\sqrt{D^2-\varepsilon^2}}{\omega-\varepsilon}$.
The coarse-grained odd-patch Green's function in VB-DMFT is instead
\begin{equation}
   G_-(\omega) = \sum_{\vk \in \mathcal{P}_{-}} \frac{1}{\omega + \mu - \ek - \Sigma_-(\omega)}.
\end{equation}
While in the 2IAM the rigid structure of the baths prevents the complete
removal of spectral weight from the chemical potential, in VB-DMFT the baths
can adjust to allow for such an effect.
When performing the VB-DMFT self-consistency loop, $\mathrm{Re} \Sigma_-(0)$ acts as a shift of the chemical potential for the odd band.
Entering the RKKY regime $\mathrm{Re} \Sigma_-(0)$ grows and can become large enough to push the
chemical potential off the band and make it insulating.
In turn, an insulating odd band enhances the intra-singlet correlations of the dimer making the solution self-consistent.

\subsection{Hybridization functions: properties of the underlying two-impurity model}

\begin{figure}[ht!]
  \includegraphics[width=8.5cm,clip=true]{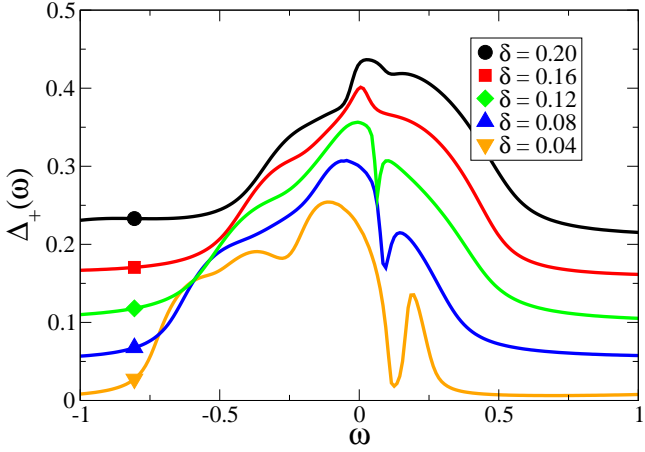}
  \caption{(Color online) $\Delta_{+}(\omega)$ for the even orbital obtained with Pad{\'e}
approximants (see Appendix~\ref{app:pade}), for various dopings at $\beta =200$.
A shift has been added between the curves for clarity.  }
  \label{fg:delta_even}
\end{figure}
%

\begin{figure}[ht!]
  \includegraphics[width=8.5cm,clip=true]{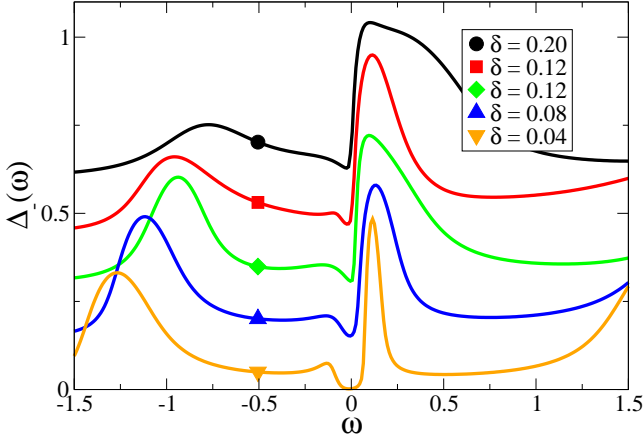}
  \caption{(Color online) $\Delta_{-}(\omega)$ for the odd orbital obtained with Pad{\'e}
approximants (see Appendix~\ref{app:pade}), for various dopings at $\beta =200$.
A shift has been added between each curves for clarity }
  \label{fg:delta_odd}
\end{figure}

To confirm the role of the VB-DMFT hybridization function in determining the
transition we analyze its behavior at different dopings.  In
Figs.~\ref{fg:delta_even} and~\ref{fg:delta_odd} we display
$\Delta_\pm(\omega)$ which appear in the self-consistent two-impurity Anderson
model solved in VB-DMFT. The hybridization function for the even orbital
$\Delta_{+}(\omega)$ shows a smooth structure with a broad peak about the
chemical potential and little variation as a function of doping.

%
The hybridization of the odd-orbital $\Delta_{-}(\omega)$
(see Fig.~\ref{fg:delta_odd}) has almost exactly the same behavior as
$A_{-}(\omega)$ up to a rescaling. At low doping, $\Delta_{-}(\omega)$ also
displays a pseudogap. Therefore, the self-consistency of the VB-DMFT equations
leads to a very non-trivial two-impurity Anderson model in the low-doping
regime: The even-orbital hybridization function is rather smooth but the
odd orbital $\Delta_{-}(\omega)$ is pseudogapped.

It is interesting to note that around the critical doping, the low-energy part
of $\Delta_{+}(\omega)$ becomes particle-hole symmetric, even though this is
not the case for $A_{+}(\omega)$. This property is most clearly seen in the
real part of $\Delta_{+} (i \omega_n)$ on the Matsubara axis (see
Fig.~\ref{fg:delta_matsubara}) that nearly vanishes for $\delta$ between $12\%$
and $16\%$.  This suggests a possible relation between the orbital-selective
transition found in VB-DMFT and the critical point of the two-impurity Kondo
model.~\cite{Jones88,affleck_conformal-field-theory_1995} The latter is known to
exist only at particle-hole symmetry, but the self-consistency can in principle
restore dynamically the symmetry and bring the system close to the critical 
point.~\cite{FerreroNutshell} These issues deserve further investigation and are
left for future work.

%

\section{Momentum-space interpolation and Fermi arcs}
\label{sec:arcs}

\subsection{Momentum-space reconstruction and comparison with larger clusters}
\label{arcs}

In this section, we address the problem of the reconstruction of
momentum-space information starting from our valence-bond description. In doing
so we will also address the reliability of our calculations by comparing our
results with those obtained with larger clusters. This will provide a 
benchmark of our approach.

An important issue in theories that use clusters to describe lattice systems is
how to infer quantities for the full lattice, starting from the information
available from the finite-cluster calculation.
In principle, this problem can be approached by a finite-size scaling study in
order to extrapolate quantities in the thermodynamic limit.
However, for cluster-DMFT theories this requires a huge computational effort,
and the size of the clusters accessible to calculations is relatively small. 

In order to obtain momentum-dependent quantities,
it is necessary to employ some form of reconstruction
based on the available cluster quantities.
Indeed, DCA methods give direct access to lattice quantities (e.g. the self-energy)
only for a few special points in the Brillouin zone: the cluster momenta.
In our simple description based on a single bond, they are $\vk=(0,0)$ and $\vk=(\pi,\pi)$.
At these momenta, the lattice quantities can be unambiguously extracted
from their cluster counterparts. For example, in our case:
$\Sigma_\text{latt}(0,0)=\Sigma_{11}+\Sigma_{12}=\Sigma_{+}$ and
$\Sigma_\text{latt}(\pi,\pi)=\Sigma_{11}-\Sigma_{12}=\Sigma_{-}$.
From the knowledge at these points, one would
like to reconstruct any point in the Brillouin zone,
using some interpolation procedure in order to avoid unphysical discontinuities
of e.g. the self-energy in momentum space.
Note that a similar procedure (reperiodization) must be used in the
real-space cluster methods like CDMFT,~\cite{kotliar_review_rmp_2006}
in order to restore the broken translation invariance.
Clearly, there is some degree of arbitrariness associated with this procedure.
The most important ingredient is the choice of the quantity to interpolate.
Since cluster quantities describe accurately the short-range
physics, it is expected that observables which are more local
(short-range) in real-space (hence less $k$-dependent)
are better suited for interpolation.

A standard method ($\Sigma\,$-interpolation) in DCA calculation
consists in interpolating the self-energy $\Sigma$ (see e.g. Ref.~\onlinecite{JarrellARPES_AkimaSpline}).
In this paper, we choose a simple interpolation, in which the lattice self-energy is given by
\begin{equation}
   \Sigma^{(\Sigma)}_\text{latt}(\vk,\omega) = \Sigma_{+}(\omega) \alpha_+(\vk) +
   \Sigma_{-}(\omega) \alpha_-(\vk),
\end{equation}
with $\alpha_\pm(\vk) = \frac{1}{2}\{1\pm\frac{1}{2}[\cos(k_x)+\cos(k_y)]\}$.
By rewriting this in terms of the on-site ($\Sigma_{11}(\omega)$)
and inter-site ($\Sigma_{12}(\omega)$) components of the cluster
self-energy
\begin{equation}
   \Sigma^{(\Sigma)}_\text{latt}(\vk,\omega) = \Sigma_{11}(\omega) +
   \frac{1}{2} \Sigma_{12}(\omega) [\cos(k_x) + \cos(k_y)].
\end{equation}
This can be viewed as a truncation of the Fourier expansion of
the lattice self-energy to the first two Fourier components. Note that,
the nearest-neighbor component of the lattice self-energy is obtained, according to
this formula, as $\Sigma_{nn}=\Sigma_{12}/4$, which is analogous to the
reperiodization procedure of CDMFT (see e.g. Ref.~\onlinecite{kotliar_review_rmp_2006}).

%
%
%

Another method ($M \,$-interpolation) has been recently introduced in
Ref.~\onlinecite{StanescuCumulantCourt,StanescuKotliarAnnPhys06} in the CDMFT method.
It consists in interpolating the cumulant,
defined as $M\equiv (\omega + \mu - \Sigma)^{-1}$.
The lattice cumulant is obtained as
\begin{equation}
   M_\text{latt}(\vk,\omega) = \alpha_+(\vk) \frac{1}{\omega+\mu-\Sigma_{+}(\omega)} +
   \alpha_-(\vk)\, \frac{1}{\omega+\mu-\Sigma_{-}(\omega)}.
\end{equation}
From $M_\text{latt}(\vk,\omega)$ it is then possible to extract a lattice self-energy by
\begin{equation}
   \Sigma^{(M)}_\text{latt}(\vk,\omega) = \omega + \mu - M_\text{latt}(\vk,\omega)^{-1}.
\end{equation}
The cumulant is the dual quantity of the self-energy in an expansion around the
atomic limit. It is a natural measure of how much the hybridization to the
self-consistent environment changes the impurity Green's function as compared
to an isolated dimer.

The $\Sigma$-interpolation is based on the assumption that the self-energy
is sufficiently short-range or small enough for all frequencies.
It corresponds to an expansion around the free-electron limit, hence it is
expected to work better at weak coupling.
On the other hand, the $M\,$-interpolation is expected to be better close to the atomic limit,
and more generally at strong coupling, for example close to a Mott insulating state where
the cumulant is more local than the self-energy.~\cite{StanescuCumulantCourt}
Other methods, like the periodization of the Green's function~\cite{KyungTremblayPeriodizationG}
have also been discussed, in the CDMFT context.
In this section, we focus on a quantitative comparison of the
$\Sigma$-interpolation  and the  $M\,$-interpolation, using a plaquette (4 sites) calculation
as a benchmark.

Let us emphasize again that for our two-site cluster the momenta
$\vk=(0,0)$ and $\vk=(\pi,\pi)$ are special. At these two points lattice quantities
are independent of the interpolation method used (since at those momenta,
one of the $\alpha$'s vanishes), while at {\em all} other
momenta the quantities reconstructed with the two methods differ.
On the other hand, in a four-site cluster (plaquette) approach, there are two
additional momenta where the description is unbiased by the interpolation
procedure, namely $\vk=(0,\pi)$ and $\vk=(\pi,0)$, which are equivalent if rotational
symmetry is not broken. Hence performing a plaquette calculation gives us the
opportunity to compare directly cluster self-energies obtained with the dimer
and the plaquette at momenta $\vk=(0,0)$ and $\vk=(\pi,\pi)$, and furthermore
provides a test for the interpolation method by comparing self-energies at
$\vk=(0,\pi)$.

\begin{figure}[ht!]
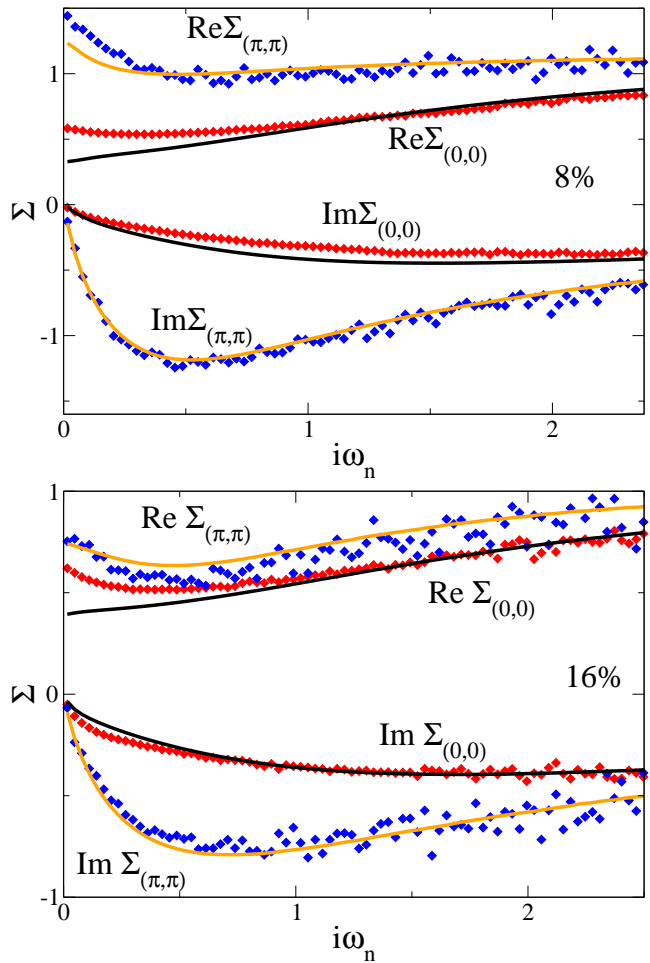

  \includegraphics[width=8.5cm,clip=true]{self_plaq_vs_dimer_08}
  \includegraphics[width=8.5cm,clip=true]{self_plaq_vs_dimer_16}
  \caption{(Color online) Real and imaginary part of dimer (solid lines) and plaquette (symbols)
self-energies at $k=(0,0)$ (black solid line and red diamond)
and $k=(\pi,\pi)$ (orange solid line and blue diamond)
for $\beta = 200$ and  $\delta= 0.08$ (upper panel) and  $\delta= 0.16$ (lower panel).
}
  \label{fg:link_vs_plaq}
\end{figure}

\begin{figure}[ht!]
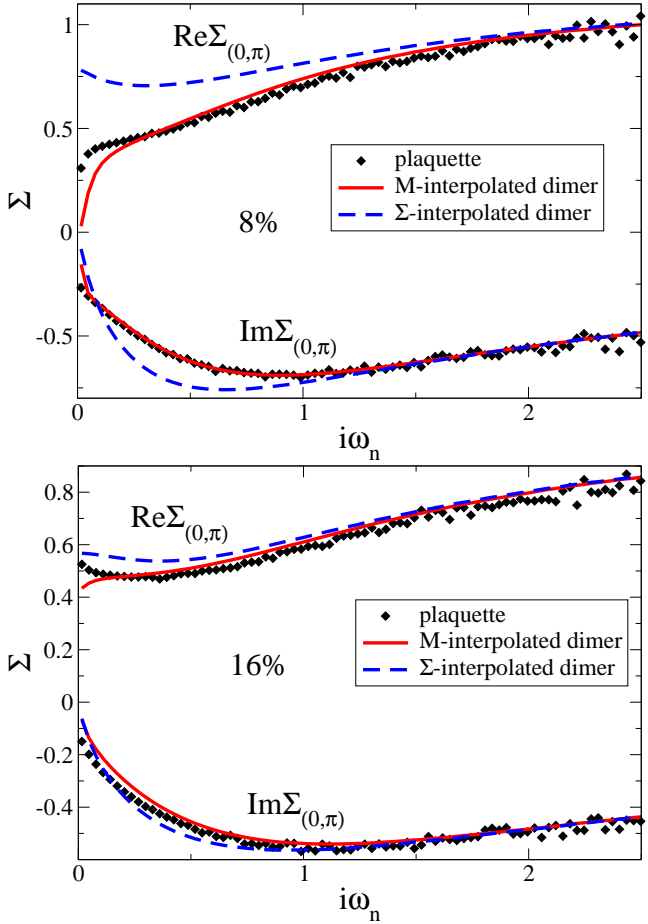

  \includegraphics[width=8.5cm,clip=true]{self_plaq_vs_dimer_reconstructed_08}
  \includegraphics[width=8.5cm,clip=true]{self_plaq_vs_dimer_reconstructed_16}
  \caption{(Color online) Comparison between the reconstructed $\Sigma_{(0,\pi)}$ using $M$-interpolation (red solid line)
and $\Sigma$-interpolation (blue dashed line) in the dimer
and the cluster self-energy of the plaquette calculation at $k=(0,\pi)$ (black diamonds),
for $\delta=0.08$ (upper panel) and $\delta=0.16$ (lower panel) . $\beta = 200$.
}
  \label{fg:reconstruct}
\end{figure}

We compare in Fig.~\ref{fg:link_vs_plaq} the results of VB-DMFT and
plaquette calculations for momenta $\vk=(0,0)$ and $\vk=(\pi,\pi)$ for
$\delta=8\%$ (upper panel) and $\delta=16\%$ (lower panel), in Matsubara frequencies.
The agreement between the two cluster calculations is good.
The descriptions of the Hubbard model given by VB-DMFT and
plaquette cluster calculations are consistent with one another for these momenta.
In order to decide which momentum-interpolation procedure is better within
VB-DMFT, we also compare in Fig.~\ref{fg:reconstruct}
(at $\delta=8\%$ in the upper panel and $\delta=16\%$  in the lower panel)
the self-energy obtained from the $\Sigma$- and $M$- interpolations, at
momentum $\vk=(0,\pi)$, to the self-energy obtained from a direct plaquette
calculation (for which $(0,\pi)$ is a cluster momentum).
%
%
Comparing the two data sets, we see that the
$M\,$-interpolation is clearly superior to the $\Sigma\,$-interpolation in
reconstructing the self-energy at $(0,\pi)$.
%

Applying the $M\,$-interpolation to the VB-DMFT results
we can qualitatively, and to a large extent quantitatively, reproduce the
larger cluster (plaquette) results, hence providing a justification to the
use of the $M$-interpolation. It is important to stress that the plaquette
cluster-momentum $\vk=(0,\pi)$ is not present as an individual orbital in the
two-site description: It is entirely reconstructed by interpolation, and as
such is the most direct test of the reconstructed momentum dependence.

\subsection{Fermi arcs and momentum differentiation}

\begin{figure}[ht!]
  \includegraphics[width=4.0cm,clip=true]{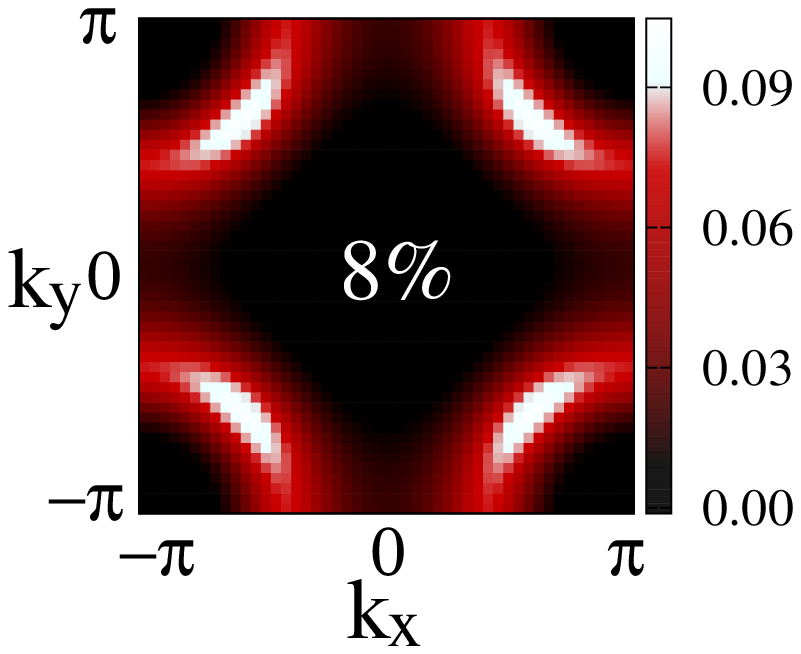}
  \includegraphics[width=4.0cm,clip=true]{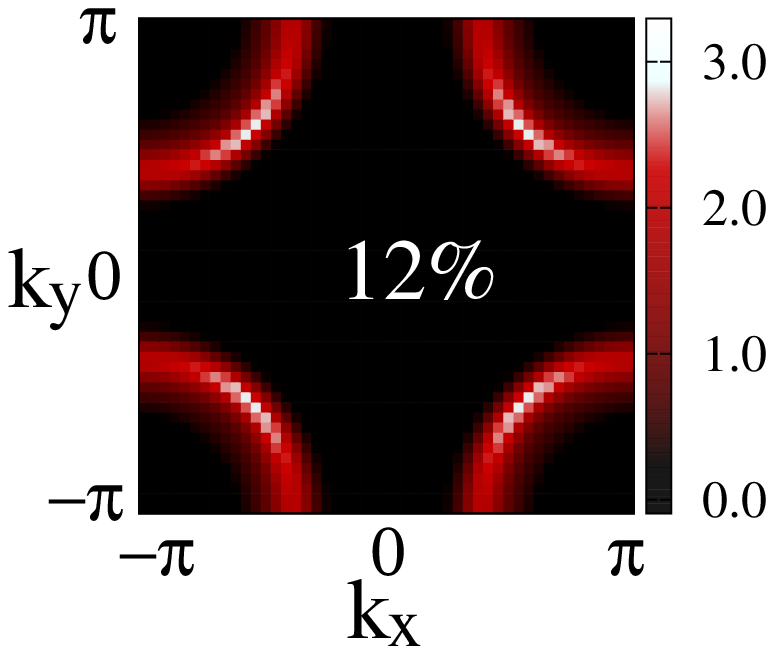}
  \includegraphics[width=4.0cm,clip=true]{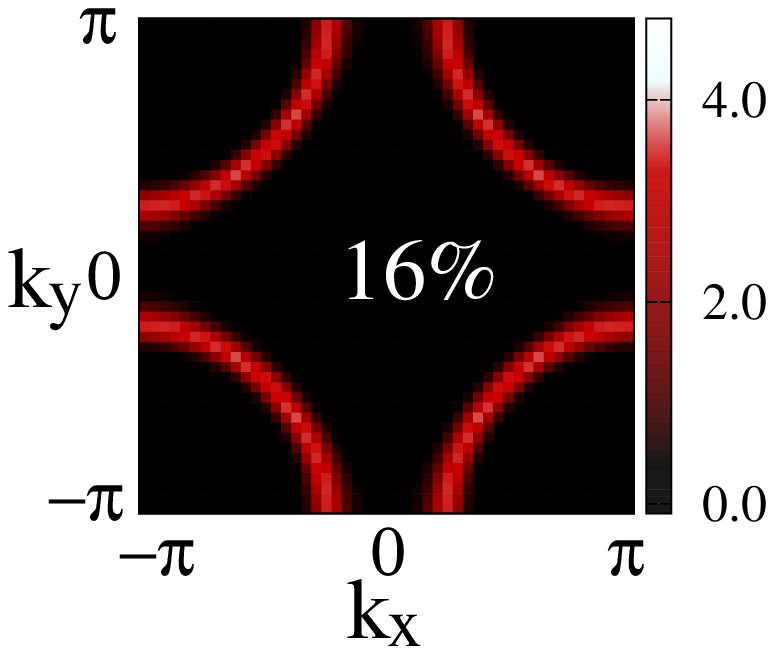}
  \includegraphics[width=4.0cm,clip=true]{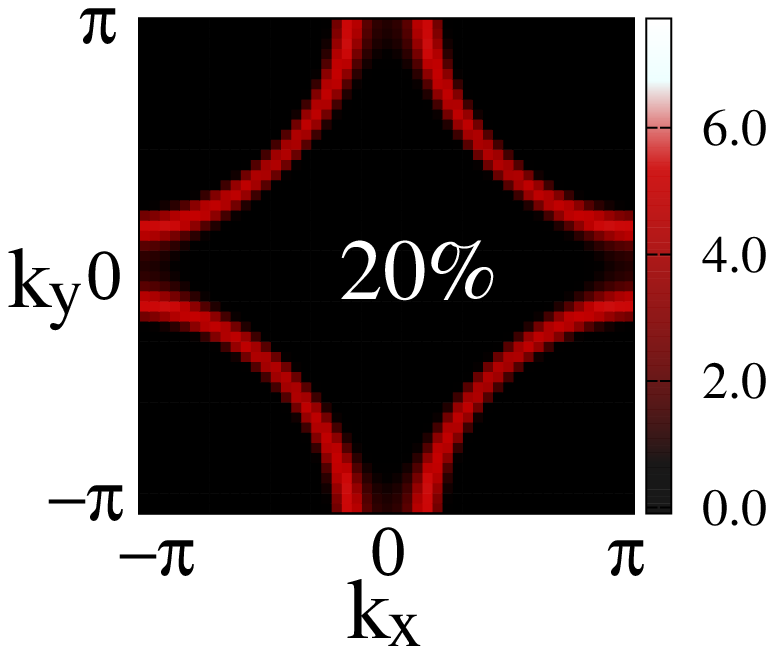}
  \caption{(Color online) Intensity maps of the spectral function $A(\vk,0)$ for different doping levels
obtained with $M$-interpolation.}
  \label{fg:A_k}
\end{figure}

We can now study momentum differentiation using the
$M\,$-interpolation.
As we shall see, VB-DMFT indeed provides a simple description of momentum
differentiation as observed in ARPES experiments.
This is illustrated by the intensity maps of the spectral function $A(\vk,0)$
displayed in Fig.~\ref{fg:A_k}.
%
%
At very high doping $\delta \gtrsim 25\%$ (not shown), cluster corrections to DMFT
are negligible and the spectral intensity is uniform along the Fermi surface.
In contrast, as the doping level is reduced, momentum differentiation sets in around the
characteristic doping at which the localization of the outer orbital takes place.
The intensity maps then display apparent ``Fermi arcs'' at finite temperature
with higher spectral intensity in the nodal direction in comparison to
antinodes,
in qualitative agreement with experiments (see e.g.
Refs.~\onlinecite{damascelli_rmp_2003,ShenARPES_Ca_Sci2005}) and earlier
CDMFT calculations with larger clusters.~\cite{parcollet_cdmft_finiteTMott_prl_2004,civelli_breakup_prl_2005,Haule2x2}
The mechanism behind the suppression of spectral weight at the antinodes at low doping
is clearly associated, in our results, to Mott localization and the importance
of singlet correlations. In technical terms, this is associated with the large real part in
$\Sigma_{-}=\Sigma(\pi,\pi)$ (cf. Fig.~\ref{fg:sigma_real}), which induces a pseudogap in the antinodal
orbital, and with the large imaginary part of
the self-energy in the $(\pi,0)$ and $(\pi,\pi)$ regions, which also contribute to
the suppression of spectral weight in the antinodal region.

\begin{figure}[ht!]
  \includegraphics[width=4.0cm,clip=true]{A_along_fermi}
  \includegraphics[width=4.0cm,clip=true]{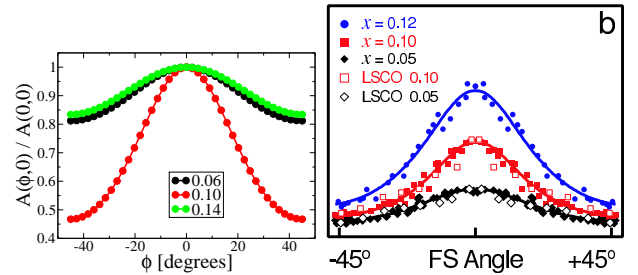}
  \caption{(Color online) Left panel: Normalized intensity
$A(\phi,0) / A(0,0)$ along the Fermi surface vs the angle to the diagonal of the Brillouin  zone in degrees
($\phi = 0$ is the node, $\phi = \pm 45$ the antinode).
The nodal intensity $A(0,0)$ is 0.045 for $\delta$=6\%,
1.66 for $\delta$=10\% and 4.61 for $\delta$=14\%. $\beta = 200$.
Right panel: Angular dependence of the spectral weight along the Fermi Surface in Ca$_{2-x}$Na$_{x}$CuO$_{2}$Cl$_{2}$
at $x = 0.05$ (black diamonds), $x = 0.10$ (red squares),
and $x =0.12$ (blue circles) along with data from La$_{2-x}$Sr$_x$CuO$_4$ for $x= 0.05$ and $x = 0.10$ (open symbols).
Figure reprinted from Ref.~\onlinecite{ShenARPES_Ca_Sci2005} (Fig.~3b). Copyright 2005 by Science. }
  \label{fg:A_along_fermi}
\end{figure}
%
In order to compare this momentum-space differentiation to experiments in a more
quantitative manner, we plot in Fig.~\ref{fg:A_along_fermi} the contrast of
the spectral intensity along the Fermi surface for different
doping levels.
This plot compares very favorably to the experimental data of
Ref.~\onlinecite{ShenARPES_Ca_Sci2005}, which are also reproduced for convenience.
In particular, we observe that the differentiation has a non-monotonous behavior,
reaching a maximum around $\delta\approx 10\%$, and then decreasing for lower dopings.

VB-DMFT is able to capture momentum differentiation reliably for
doping levels between $10\%$ and $20\%$.
For very low doping ($\delta \lesssim 8\%$) the $M$-interpolated self-energy
develops singularities on lines in momentum space, leading to lines of zeroes
of the Green's function and to the breakup of the Fermi
surface.~\cite{StanescuCumulantCourt,Essler02,Dzyaloshinski03,Berthod06,Yang06}
It would be very instructive to relate this with recent quantum oscillation
experiments.~\cite{doiron_nature} However, in this regime, a better momentum
resolution (larger clusters) is necessary to obtain reliable results.

\section{Conclusion}
\label{sec:conclusion}

In this paper, we presented in detail the valence-bond DMFT approach
to correlated electrons~\cite{VBDMFT} and used it to treat the two-dimensional
Hubbard model with nearest-neighbor and next-nearest neighbor hopping.
The approach reduces to single-site DMFT when intersite correlations
are unimportant. This is the case at large doping levels.
Near the Mott transition, at lower doping levels, these correlations dominate
the physics and lead to the phenomenon of momentum-space differentiation.

This phenomenon corresponds to the destruction of coherent quasiparticle
excitations in the antinodal regions of the Brillouin zone. In those
regions, a pseudogap opens and quasiparticles become increasingly incoherent
as the doping level is reduced. In contrast, in nodal regions, quasiparticles
are protected and their lifetime actually increases as the pseudogap opens.
The physics of the low-doping regime is dominated by strong singlet
correlations between nearest-neighbor sites.

VB-DMFT is a minimal cluster description of a low-dimensional
strongly-correlated system in terms of two effective degrees of
freedom, associated to each of the important regions in momentum space
(nodal and antinodal). These two degrees of freedom are treated as the two
orbitals of an effective dimer impurity model. Momentum-space differentiation
emerges as an orbital-selective Mott transition in which a pseudogap opens in the
spectrum of the antinodal degree of freedom, while the nodal one remains a
coherent Fermi liquid.
The simplicity of the approach allows for a highly accurate numerical solution
of the VB-DMFT equations. It also allows for the use of a semi-analytical technique,
the rotationally-invariant slave-bosons method,
as  an approximate impurity solver.
%

Comparisons of the results of  VB-DMFT and plaquette calculations
put the cluster extensions of DMFT on much firmer footing. It has been known for a while
that in the high-temperature and high-enough doping regime, single-site DMFT
is very accurate and cluster corrections are quantitatively small.
On the other hand, the validity of cluster extensions of DMFT in the underdoped regime,
where momentum-space differentiation is strong, is not as universally accepted.
Consistency of the results between 2-site calculations and
calculations with larger clusters
provides support to the validity of cluster approaches.

The qualitative picture that emerges from VB-DMFT is in excellent
qualitative agreement with photoemission results in the normal state of the
copper-oxide based high-temperature superconductors.
The selective destruction of quasiparticles at the antinodes is
associated with the `Fermi arcs' observed in ARPES.
Comparison of the evolution of the tunneling density of states with temperature
against experimental data is also encouraging. Many more detailed
comparisons of VB-DMFT against various other spectroscopies should
be carried out in future work in order to determine the strengths and the limitations
of the method and in order to further advance our understanding and our ability
to capture with simple models some of the physical properties of cuprates.


\acknowledgments
We thank C.~Berthod, J.~C.~Campuzano, L.~de'~Medici, J.~C.~Seamus~Davis,
M.~Fabrizio, K.~Haule, F.~Lechermann, A.~J.~Millis, T.~M.~Rice, A.~Sacuto and
P.~Werner for useful discussions.  We also thank F. Assaad and M. Aichhorn for
providing results using the stochastic maximum entropy method.  We acknowledge
support from The Partner University Fund (PUF-FACE), ICAM, the ANR (under
grants ETSF, GASCOR and ECCE) and the NSF-Materials World Network.  One of us (PSC)
thanks CPHT and IPhT-Saclay for hospitality.

\appendix

\section{DCA with Brillouin-zone patches of arbitrary shapes}
\label{Appendix-dca}

In this appendix, we review for completeness
the basic formalism of DCA,~\cite{DCA_98,DCA_long2000,maier_cluster_rmp_2005}
and discuss in particular its extension to Brillouin-zone patches with equal volume and arbitrary shape.

DCA can be seen as an approximation to the Luttinger-Ward (LW) functional of a
lattice theory. In the LW functional the conservation of
momentum at the vertex of diagrams is accounted for by the function
\begin{equation}
   \Delta (\vk_1, \vk_2, \vk_3, \vk_4) =
      N \delta_{\vk_1+\vk_2,\vk_3+\vk_4}.
\end{equation}
In single-site DMFT momentum conservation at the internal vertices of the diagrams is
ignored and $\Delta \equiv 1$.
DCA attempts at partially restoring momentum conservation by partitioning the Brillouin
zone into $N_p$ patches ${\mathcal P}_\vK$ centered around a subgroup of
$N_p$ momenta $\vK$ and approximating the momentum conservation with
\begin{equation}
   \Delta (\vk_1, \vk_2, \vk_3, \vk_4) \sim
      N_p \delta_{\vK_1+\vK_2,\vK_3+\vK_4},
\end{equation}
where $\vK_i$ is the representative vector of the patch containing
$\vk_i$. This corresponds to taking into account momentum conservation among
the patches and discarding momentum conservation inside the single patch.

The DCA LW functional contains the same diagrams as the original
lattice functional with all the internal Green's functions replaced by the
coarse-grained Green's functions
\begin{equation}
   G(\vK) = \frac{1}{N_K} \sum_{\vk \in {\mathcal P}_\vK} G_\mathrm{latt}(\vk),
\end{equation}
where $N_K$ is the number of momenta contained in the patch ${\mathcal P}_\vK$
(the volume of the patch).
To see this we can consider the simplest graph contributing to the LW
functional. The contribution of this graph to the lattice
functional is given by
\begin{eqnarray}
   \Phi_\mathrm{latt} &=& \frac{1}{N^4} \sum_{\vk_1\vk_2\vk_3\vk_4}
   G_\mathrm{latt}(\vk_1) G_\mathrm{latt}(\vk_2) G_\mathrm{latt}(\vk_3) G_\mathrm{latt}(\vk_4)
   \nonumber \\
   && U^2 N^2 \delta_{\vk_1+\vk_2,\vk_3+\vk_4}.
\end{eqnarray}
Replacing the original momentum conservation with the DCA approximation we
obtain
\begin{eqnarray}
   \Phi_\mathrm{DCA} &=& \frac{1}{N^4} \sum_{\vK_i} \sum_{\tilde{\vk}_i}
   G_\mathrm{latt}(\vK_1+\tilde{\vk}_1) G_\mathrm{latt}(\vK_2+\tilde{\vk}_2)
   \nonumber \\ && \times
   G_\mathrm{latt}(\vK_3+\tilde{\vk}_3) G_\mathrm{latt}(\vK_4+\tilde{\vk}_4)
   \nonumber \\
   && U^2 N_p^2 \delta_{\vK_1+\vK_2,\vK_3+\vK_4}
   \nonumber \\
   &=& \sum_{\vK_i} \frac{ N_{K_1} N_{K_2} N_{K_3} N_{K_4} }{N^4}
   \nonumber \\
   &&
   \left( \frac{1}{N_{K_1}} \sum_{\tilde{\vk}_1}
   G_\mathrm{latt}(\vK_1+\tilde{\vk}_1) \right) \times
   \nonumber \\ &&
   \times \ldots \times
   \left( \frac{1}{N_{K_4}} \sum_{\tilde{\vk}_4}
   G_\mathrm{latt}(\vK_4+\tilde{\vk}_4) \right)
   \nonumber \\
   && U^2 N_p^2 \delta_{\vK_1+\vK_2,\vK_3+\vK_4}
   \nonumber \\
   &=& \sum_{\vK_i} \frac{ N_{K_1} N_{K_2} N_{K_3} N_{K_4} }{N^4}
   G(\vK_1) \ldots G(\vK_4)
   \nonumber \\
   && N_p^2 U^2 \delta_{\vK_1+\vK_2,\vK_3+\vK_4}.
\end{eqnarray}
It is then clear that if the number of $\vk$- points (the BZ volume)
is the same for every patch, the prefactor can be simplified, yielding
\begin{eqnarray}
   \Phi_\mathrm{DCA} &=& \frac{ 1 }{N_p^4} \sum_{\vK_i}
   G(\vK_1) \ldots G(\vK_4)
   \nonumber \\
   && N_p^2 U^2 \delta_{\vK_1+\vK_2,\vK_3+\vK_4}.
\end{eqnarray}

This functional corresponds to the functional of a problem with $N_p$ momenta
and hence can be obtained by the solution of a cluster impurity problem.
The crucial observation
is that, as long as the volume is the same for all the patches, the DCA
functional is the functional of a cluster problem which, once expressed in real
space coordinates, retains purely local interactions, precisely identical to those of
the original Hubbard
model. This ensures that this procedure does not generate additional
interactions in the cluster.

However, the {\it shape} of the patches is not constrained in this
procedure. A possible route to exploit this freedom is to notice that DCA can
also be interpreted as an approximation to the lattice self-energy (see Sec.~\ref{sec:model}). Indeed DCA
corresponds to approximating the self-energy by a constant value in each patch
\begin{equation}
   \Sigma (\vK_i + \tilde{\vk}, \omega) \sim
   \Sigma (\vK_i, \omega) \qquad \forall \, \vK_i+\tilde{\vk}
   \in {\mathcal P}_{\vK_i}.
\end{equation}
Hence using physical intuition patches can be shaped in such a way to enclose
regions with definite properties, such as nodal and antinodal regions for example.


\section{Analytical continuations using Pad{\'e} approximants}
\label{app:pade}

In order to perform the analytical continuations shown earlier in this
paper, we have used $N$-point Pad{\'e} approximants~\cite{vidberg_pade}
\begin{equation}
  C_N(z) = A_N(z) / B_N (z),
\end{equation}
where $A_N$ and $B_N$ are polynomials of order $(N-1)/2$ and $(N-1)/2$ if $N$
is odd and $(N-2)/2$ and $N/2$ if $N$ is even. The Pad{\'e} approximant
$C_N$ can alternatively be written as a continued
fraction
\begin{equation}
  C_N(z) = \frac{a_1}{1+} \frac{a_2 (z-z_1)}{1+} \ldots \frac{a_N (z-z_{N-1})}{1}.
\end{equation}
The polynomials $A_N$ and $B_N$ are then given by a recursion formula
\begin{eqnarray}
  A_{n+1} (z) = A_n (z) + (z-z_n) a_{n+1} A_{n-1} (z) \\
  B_{n+1} (z) = B_n (z) + (z-z_n) a_{n+1} B_{n-1} (z),
\end{eqnarray}
with
\begin{equation}
  A_0 = 0, \qquad A_1 = a_1, \qquad B_0 = B_1 = 1.
\end{equation}
In order to construct the complex function $C_N(z)$, we impose that
it is equal to the function $f$ to be continued on the real
axis at the first $N$ Matsubara frequencies
\begin{equation}
  C_N (i \omega_n) = f (i \omega_n) \quad \forall \quad n = 1, \ldots, N.
\end{equation}
This constraint can be achieved by determining the coefficients
$a_n$ with the recursion
\begin{eqnarray}
  a_n = g_{n,n}, \quad g_{1,n} = f(i \omega_n), \quad n = 1, \ldots, N \\
  g_{p,q} = \frac{g_{p-1,p-1} - g_{p-1,q}}{(i \omega_q - i \omega_{p-1}) g_{p-1,q}}.
\end{eqnarray}

In this paper, the Monte Carlo data on the Matsubara axis has been averaged
over $5 \times 10^8$ measures. A Pad{\'e} approximant was computed for
$\Sigma_\pm$ imposing that they match over the first 200 Matsubara frequencies
for a temperature $1/\beta = 1/200$. The self-energies $\Sigma_\pm(\omega)$
obtained on the real-frequency axis using this procedure were then used to
compute the spectral functions through
\begin{equation}
  A_\pm(\omega) = -\frac{1}{\pi} \mathrm{Im} \sum_{\vk \in \mathcal{P}_\pm}
    \frac{1}{\omega + \mu - \ek - \Sigma_\pm(\omega)}.
\end{equation}

\begin{figure}[ht!]
  \includegraphics[width=8.5cm,clip=true]{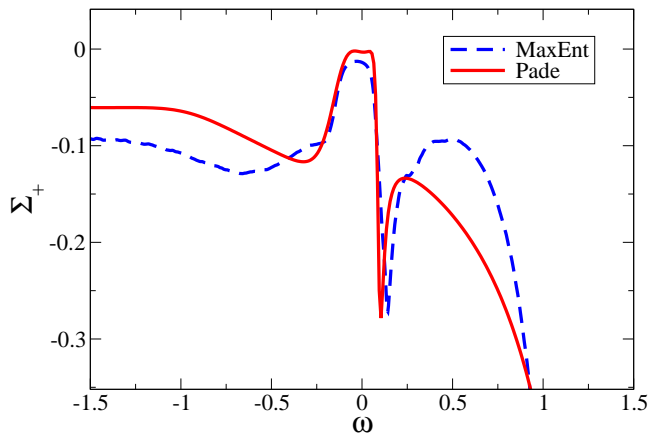}
  \caption{(Color online) Even self-energy on the real frequency axis as obtained
  using Pad{\'e} approximants (solid line) and by maximum entropy (dashed line).
  $\delta = 8\%$ and $\beta = 200$.}
  \label{fg:maxent_vs_pade}
\end{figure}

In order to test the quality of the analytical continuations, we constructed
a Pad{\'e} approximant for several independent runs checking that they lead
to qualitatively similar results. Indeed, spurious poles sometimes appear
on the real-frequency axis producing unphysical results.
We also compared the Pad{\'e} approximant with the outcome of a stochastic maximum
entropy method.~\cite{beach_maxent,sandvik_maxent} A typical outcome is
shown in Fig.~\ref{fg:maxent_vs_pade}. Even if both method lead to results
that are slightly different at a quantitative level, they display the same main
qualitative features. Therefore all our physical conclusions stated earlier
do not depend on the analytical continuation method.

\section{Raw CTQMC data}
\label{app:Matsubara}

For completeness, we display the raw CTQMC data on the Matsubara axis for the
Green's functions $G_\pm$, the self-energies $\Sigma_\pm$ and the hybridization
functions $\Delta_\pm$. The analytical continuations using Pad{\'e} approximants
(see Appendix~\ref{app:pade}) have been performed on this data.

\begin{figure*}[ht!]
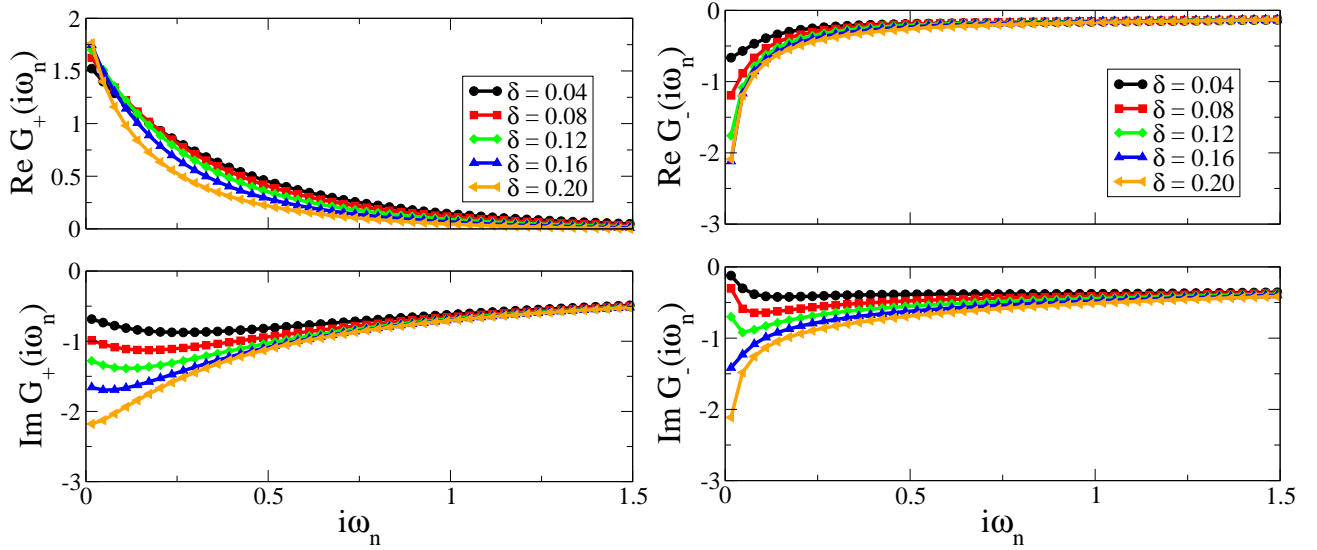

  \includegraphics[width=8.5cm,clip=true]{green_mat_even}
  \includegraphics[width=8.5cm,clip=true]{green_mat_odd}
  \caption{(Color online) Even and odd Green's functions $G_\pm(i \omega_n)$ on the Matsubara
  axis for various dopings at $\beta = 200$.}
  \label{fg:green_matsubara}
\end{figure*}

\begin{figure*}[ht!]
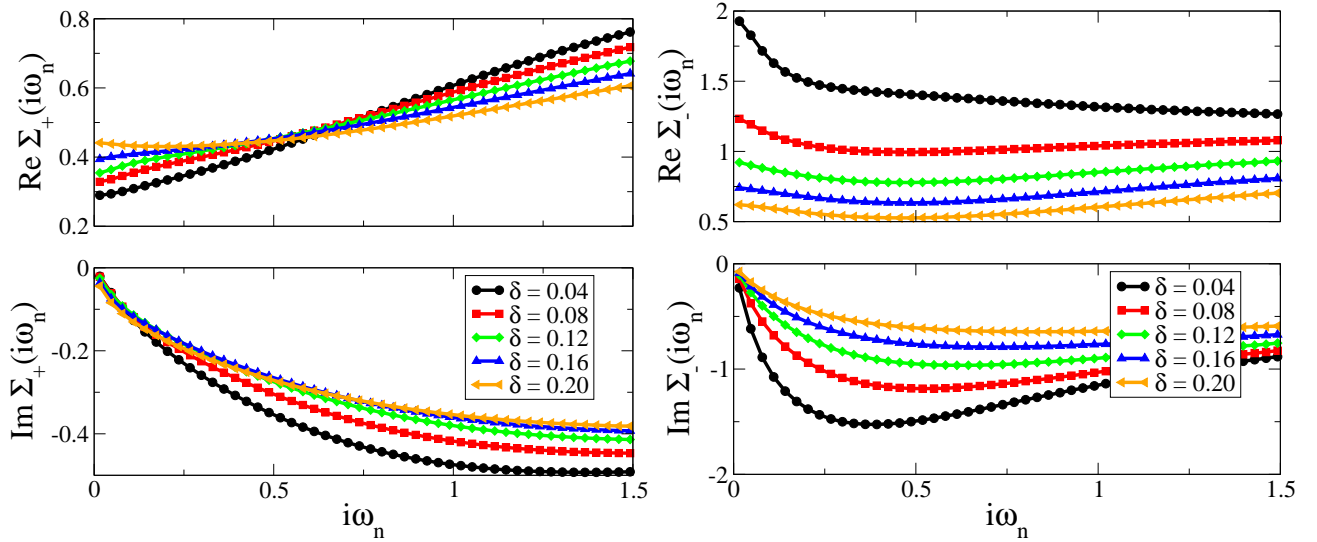

  \includegraphics[width=8.5cm,clip=true]{self_mat_even}
  \includegraphics[width=8.5cm,clip=true]{self_mat_odd}
  \caption{(Color online) Even and odd self-energies $\Sigma_\pm(i \omega_n)$ on the Matsubara
  axis for various dopings at $\beta = 200$.}
  \label{fg:self_matsubara}
\end{figure*}

\begin{figure*}[ht!]
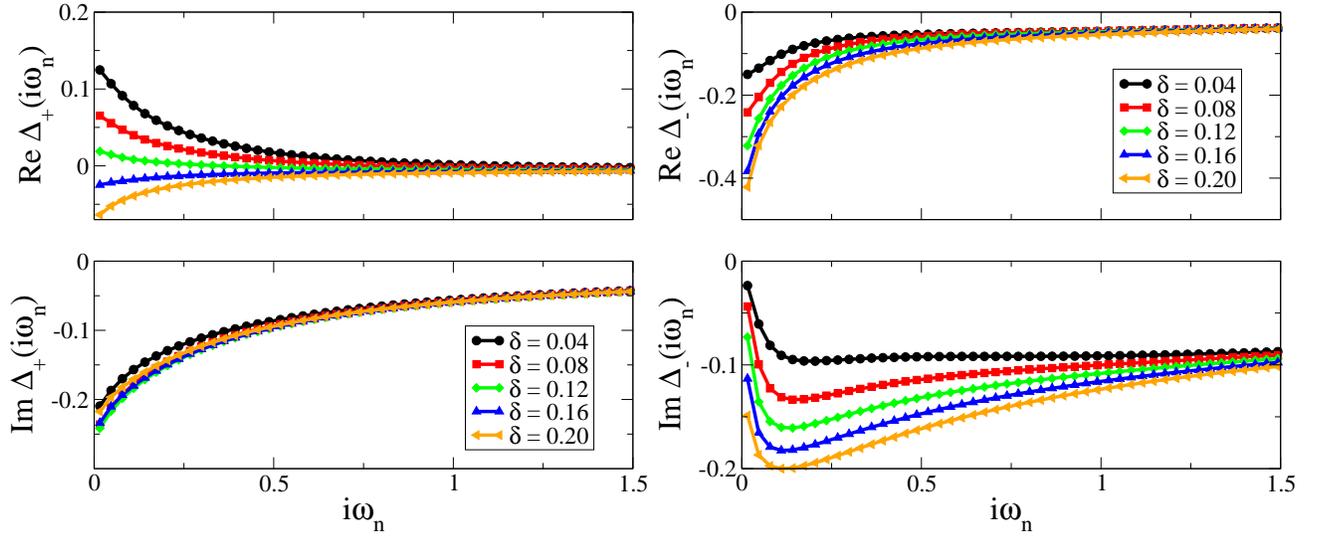

  \includegraphics[width=8.5cm,clip=true]{delta_mat_even}
  \includegraphics[width=8.5cm,clip=true]{delta_mat_odd}
  \caption{(Color online) Even and odd hybridization functions $\Delta_\pm(i \omega_n)$ on the Matsubara
  axis for various dopings at $\beta = 200$.}
  \label{fg:delta_matsubara}
\end{figure*}

\end{document}